\documentclass{aa}  
\usepackage{graphicx}
\usepackage{txfonts,textcomp}
\usepackage{booktabs}
\usepackage{float}
\usepackage{tabularray}
\usepackage{hyperref}
\usepackage{mhchem} 
\usepackage{multirow}
\usepackage{xcolor}
\usepackage{amsmath}

\begin{document}

\title{The impact of surface acetylene cyclotrimerization on the abundance of aromatic hydrocarbons in carbon-rich asymptotic giant branch stars}
\titlerunning{Impact of acetylene cyclotrimerization on aromatics}
\authorrunning{M. S. Murga et al.}
\author{M. S. Murga
    \inst{1}
          \and
          I. V. Loginov \inst{1,2}
          \and
          D. S. Wiebe \inst{1,2}
          \and
          D. R. Fedotova\inst{3}
          \and
          V. S. Krasnoukhov \inst{4,5}
          \and
          I. O. Antonov \inst{4}
          }

   \institute{Institute of Astronomy of Russian Academy of Sciences,
              Pyatnitskaya str. 48, 119017, Moscow, Russia\\
              \email{murga@inasan.ru}
         \and
             Faculty of Chemistry, Lomonosov Moscow State University, Universitetsky pr. 13, 119234, Moscow, Russia 
             \and Skolkovo Institute of Science and Technology, Bolshoy Boulevard, 30c1, 121205, Moscow, Russia 
        \and
             Samara Branch of P. N. Lebedev Physical Institute of the Russian Academy of Sciences, Novo-Sadovaya str. 221, 443011, Samara, Russia 
        \and
            Ural Federal University, 19 Mira str., 620062, Ekaterinburg, Russia
             }

\date{Received ; accepted }

\abstract
{}
{This work investigates the catalytic role of dust grains in forming aromatic hydrocarbons via acetylene cyclotrimerization on their surfaces within the circumstellar envelopes of carbon-rich asymptotic giant branch (AGB) stars.}
{We present a comprehensive computational astrochemical model coupling the gas-phase, gas-surface, and surface  (cyclotrimerization) reactions, and the physical evolution of the dust grains (coagulation). The model expands upon the basic chemical network from previous models, enhancing them with updated reactions involving hydrocarbons up to pyrene. We applied this model to simulate the chemical evolution of the envelope of the prototypical AGB star IRC+10216, utilizing physical conditions derived from a hydrodynamical model available in literature. To quantify the impact of surface chemistry, we compared scenarios with and without the cyclotrimerization reaction, further testing the sensitivity of our results by varying the key parameter of hydrocarbon desorption energy.}
{We find that surface-catalyzed cyclotrimerization is a viable pathway for aromatic formation in circumstellar environments, capable of enhancing the total abundance of aromatic species by up to an order of magnitude. Crucially, we show that gas-phase chemistry and dust surface processes are intrinsically linked; their synergistic evolution should be modeled self-consistently to accurately predict chemical abundances. This work underscores that constraining uncertain parameters, particularly desorption energies of hydrocarbons, is essential for future realistic modeling of astrochemical processes in evolved stellar systems.}
{}

\maketitle

\section{Introduction} \label{sec:intro}

The formation of polycyclic aromatic hydrocarbons (PAHs) in astrophysically favorable environments -- particularly in circumstellar envelopes of asymptotic giant branch (AGB) stars -- has been extensively studied over several decades~\citep{frenklach89, cherchneff92, cadwell94, cherchneff12}. Nevertheless, a complete understanding of the evolution of aromatic hydrocarbons remains elusive.

Acetylene (\ce{C2H2}) has been identified as a key molecule in PAH and carbon dust formation in AGB stars~\citep{gail87, cherchneff12, dhanoa14, pentsak24}. The pioneering kinetic model by \citet{frenklach89}, based on combustion chemistry networks, introduced the hydrogen abstraction–acetylene addition (HACA) mechanism as a primary pathway for PAH growth. While this work established chemical routes for PAH formation under circumstellar conditions, it also revealed that the process is efficient only within a narrow gas temperature range ($\approx 900-1000$~K) and in stars with high mass-loss rates. These findings prompted further exploration of alternative PAH formation pathways and detailed reaction analysis. 
  
Alternative mechanisms, such as the recombination of two propargyl radicals proposed by \citet{miller90} and \citet{cherchneff92}, proved more promising in circumstellar environments. However, stationary-flow models alone could not account for the observed PAH abundance~\citep{cherchneff92, helling96}. By coupling astrochemical kinetical model with envelope dynamics -- including shock waves from stellar pulsations -- \citet{willacy98, cherchneff12} significantly enhanced PAH production efficiency. Their simplified estimates yielded a carbonaceous dust-to-gas mass ratio of $\approx 1.2 \cdot 10^{-3}-5.8 \cdot 10^{-3}$, consistent with observational and galactic dust evolution values of $10^{-4}-10^{-3}$~\citep{ferrarotti06, zhukovska08}.

While most studies focus on gas-phase PAH formation, concurrent dust nucleation and growth processes are equally critical. Carbonaceous dust nucleation in AGB stars may occur via homogeneous or heterogeneous pathways~\citep{gail88, cadwell94}. \citet{cadwell94} proposed induced nucleation, with silicon carbide (SiC) particles serving as nucleation seeds due to their higher formation temperatures~\citep{frenklach89_sic, lorenz94, kozasa96}. These particles, observed in AGB stars and meteorites~\citep{speck09}, provide substrates for PAH and other hydrocarbons adsorption, making heterogeneous nucleation far more effective than homogeneous processes. Notably, \citet{cadwell94} hypothesized that reactions on the SiC surface could explain the high PAH abundance in the interstellar medium (ISM). The PAHs formed on grains might later desorb under ultraviolet (UV) radiation during post-AGB or planetary nebula phases.

Apart from providing nucleation sites, SiC surfaces catalyze aromatic formation. Studies confirm efficient acetylene chemisorption on SiC active sites~\citep{frenklach89_sic, carmer93, cho01, ruiz13, pollmann14}, followed by cyclotrimerization (CTM) to benzene (\ce{C6H6}) -- a process experimentally validated by \citet{zhao16}. Subsequent surface reactions could coat SiC particles with graphene-like layers. In its turn, \citet{merino14} suggested that SiC surfaces first graphitize, then fragment into PAHs via hydrogen passivation or etching. Both \citet{zhao16} and \citet{merino14} underscore SiC importance in PAH formation.

Carbonaceous materials also facilitate acetylene CTM, a key carbocatalytic process~\citep{pentsak20, pentsak24}. First discovered by \citet{berthelot1867} and then independently by \citet{zelinsky1923}, this reaction gained renewed interest through density functional theory (DFT) studies by \citet{gordeev20}, who demonstrated its viability on zigzag PAH edges with low activation barriers. \citet{gordeev20} studied several PAHs containing from 6 to 361 C atoms and concluded that most of them (except phenyl radical) can be considered as catalysts for acetylene CTM which demonstrates the versatility of this mechanism.

In this work, we quantify the impact of acetylene CTM on SiC and PAH zigzag edges in AGB envelopes. Combining the dynamical model of IRC+10216 from \citet{cherchneff12} with an astrochemical model -- incorporating gas-phase, surface reactions, and dust growth (adsorption and coagulation) -- we assess PAH abundance enhancement under realistic circumstellar conditions.

\section{The astrochemical model}

\subsection{Main statements}

 The model used in this work builds upon our previous study, \citet[hereafter Paper~I]{murga24}, in which we incorporated acetylene CTM on SiC particle surfaces into the chemical reaction network. The key assumptions of the model in Paper~I are as follows:
 
\begin{itemize}

\item The physical conditions for astrochemical calculations -- gas number density ($n_{\rm gas}$) and temperature ($T_{\rm gas}$) -- are adopted from \cite{cherchneff12, willacy98}. These conditions were derived from a dynamical model of the envelope of the star IRC+10216, accounting for periodic shock wave propagation related to stellar thermal pulses. Such propagation results in a sharp increase in the gas density and temperature. Cooling of disturbed gas occurs in two steps due to: 1) dissociation of H$_2$ molecules; 2) adiabatic expansion. After cooling, gas moves as a stationary flow before being disturbed by another shock wave.  

\item SiC particles grow via SiC molecule adsorption, while coagulation is neglected. The initial number of particle seeds is set to $10^{-12}$ relative to the number of H$_2$ molecules.
\item The chemical reaction network is based on that from \citet{cherchneff12}. The original network consists of 377 gas-phase reactions, with benzene as the largest molecule. In Paper~I, this network has been extended with surface reactions involving acetylene adsorption on SiC particles and subsequent CTM into benzene.
\end{itemize}

In the present work, we present the updated model (designated as BRAHMA), which further enhances the previous one by incorporating the following processes:

\begin{itemize}
\item Chemisorption of acetylene molecules on PAH zigzag edges and their CTM;
\item Aromatic growth pathways from benzene to pyrene;
\item Dust grain growth via hydrocarbon adsorption;
\item Dust grain coagulation.
\end{itemize}

We exclude homogeneous nucleation of SiC particles, as the results of Paper~I suggest that SiC dust forms predominantly close to the star before aromatic molecules and carbonaceous particles emerge. Subsequent dust growth is assumed to occur via hydrocarbon adsorption onto SiC particles, consistent with the induced nucleation mechanism studied in \citet{frenklach89_sic, cadwell94}. Homogeneous nucleation and growth of purely carbonaceous particles are omitted, as test calculations indicate their inefficiency compared to heterogeneous nucleation, while their inclusion significantly slows down computations. 

Our reaction network is limited to pyrene, justified by experimental and theoretical considerations. Studies on the formation of interstellar dust analogs show that small PAHs are predominantly present in gas during the growth process \citep{santoro20}. This is partially due to observational constraints but it also reflects the lower abundance of larger PAHs. Similarly, combustion theories suggest that moderate-sized PAHs dominate soot formation \citep{WANG11, jacobson20}. Thus, we argue that restricting PAH size to pyrene is well founded. Intriguingly, PAHs in the experiment on the formation of dust analogs~\citep{santoro20} often bear aliphatic substituents (e.g., \ce{-CH2}, \ce{-C2H3}), which may bridge aromatic units, yielding hydrogen-rich, irregular structures. Such PAHs are present in our network and also participate in the dust growth process. 

\subsection{Kinetic equations}

The astrochemical model presented above is described by the system of discrete kinetic equations, which include separate equations for the number density of species in gas ($n_i$), on the dust surface ($n_i^{\rm s}$), and dust grains themselves ($n_i^{\rm d}$). Kinetic equations for the number densities normalized to $n_{\rm gas}$ are
\begin{eqnarray}
\label{maineq}
\frac{dn_i}{dt} = && \sum\limits_{j} k_j n_j + \sum\limits_{l,m} k_{lm} n_l n_m n_{\rm gas}  + \sum\limits_{o,p,t} k_{opt} n_o n_p n_t n_{\rm gas}^2 \nonumber \\
&& - n_i \left[ \sum\limits_{s} k_s +  \sum\limits_{q} k_{iq} n_q n_{\rm gas} +  \sum\limits_{v,w} k_{ivw} n_v n_w n_{\rm gas}^2\right]  \nonumber \\
&& + \sum_{z=1}^{N_a}k^{\rm des}_{iz} n_{iz}^s -\sum_{z=1}^{N_a}k^{\rm ad}_{iz} n_i,\\
\frac{dn_{iz}^{\rm s}}{dt} = && k^{\rm ad}_{iz}n_i - k^{\rm des}_{iz}n_{iz}^{\rm s} + k^{\rm sr}_{ijz}n_{jz}^{\rm s}n_{\ce{C2H2}}n_{\rm gas}-k^{\rm sr}_{liz}n_{iz}^{\rm s}n_{\ce{C2H2}}n_{\rm gas},\\
\label{maineq3}
\frac{dn_i^{\rm d}}{dt} = &&\frac{1}{2} \sum\limits_{j=1}^{N_a} \sum\limits_{l=1}^{N_a} C_{jli}\beta_{jl} n_{j}^{\rm d} n_{l}^{\rm d} n_{\rm gas}  - n_{i}^{\rm d}\sum\limits_{j=1}^{N_{a}} \beta_{ij} n_j^{\rm d} n_{\rm gas} 
\end{eqnarray}
where $k_j$ and $k_s$, $k_{lm}$ and $k_{iq}$, $k_{opt}$ and $k_{ivw}$ are rate coefficients of unimolecular, bimolecular and termolecular reactions, correspondingly, $k^{\rm des}_{iz}$ and $k^{\rm ad}_{iz}$ are desorption and adsorption rates, $k^{\rm sr}_{ijz}$ (and $k^{\rm sr}_{liz}$) is a rate coefficient of a surface reaction between \ce{C2H2} and $j$($i$)-th molecule on the $z$-th grain with formation of the $i$($l$)-th molecule. The second index in $k^{\rm sr}_{ijz}$ ($j$) can correspond to sC$_2$H$_2$\footnote{Hereafter, the prefix `s' means that the molecule is on surface} and sC$_4$H$_4$, the first index in $k^{\rm sr}_{liz}$ ($l$) --- to sC$_4$H$_4$ or sC$_6$H$_6$. $C_{jli}$ is the value related to the mass ($m_{[{\rm index}]}$) that was put to the $i$-th bin after coagulation of $j$-th and $l$-th grains, which we calculated according to \citet{akimkin15} as
\begin{equation}
    C_{jli} = \begin{cases}
    \epsilon, & \text{if $m_i= {\rm max}\{m_n\} \; {\rm where }\; m_{n}<m_j+m_l $},\\
    1-\epsilon, & \text{if $m_i= {\rm min}\{m_n\} \; {\rm where }\; m_{n}>m_j+m_l $},\\
    0,  & \text{other cases}.
  \end{cases}
\end{equation}

In Eq.~\ref{maineq3}, $N_{a}$ is the number of dust grain populations, the factor $\beta_{jl}$ is the frequency of collisions between $j$-th and $l$-th grains, which is calculated as
\begin{equation}
    \beta_{jl} = 2.2\sqrt{\frac{\pi k_{\rm B} T_{\rm gas}}{2\mu_{jl}}}(2a_{j}+2a_{l})^2, 
\end{equation}
where $k_{\rm B}$ is the Boltzman constant, $\mu_{jl}$ is the reduced mass of $j$-th and $l$-th grains, $a_j$ and $a_l$ are their radii, and factor 2.2 is the Van der Waals enhancement factor~\citep{harris88}.

At each calculation step we performed two corrections. The first one was to take into account the change of the grain mass and radius due to surface reactions. For the $i$-th grain, we calculated the difference, $dm_i$, between masses of all surface components at the current and previous steps and added this difference to the mass of the $i$-th grain. The radius of the $i$-th grain, $a_i$, was recalculated from the new mass.

The second correction was to redistribute surface components between grains according to changes in dust grain number densities after coagulation process for the time interval $dt$. To find the corrected number density of $i$-th surface component on the $z$-th dust grain $\left( n^{\rm s}_{iz}\right)^{\rm corr}$, we modified Eq.~\ref{maineq3}. Instead of the dust grain number density, the equation was rewritten for the number density of surface components. The first term in the right-hand side of Eq.~\ref{maineq3} was multiplied by the sum of number densities of surface components of each grains normalized to the number densities of these grains. The number density of the $z$-th dust grains was replaced by the uncorrected number density of the $i$-th component on the $z$-th dust grain in the second term in the right part of Eq.~\ref{maineq3}. Since we calculated the number density instead of its change, the right part was multiplied by $dt$. The obtained expression for $\left( n^{\rm s}_{iz}\right)^{\rm corr}$ is
\begin{align}
\left(n^{\rm s}_{iz}\right)^{\rm corr}  = dt \times \Bigg( &\frac{1}{2} \sum\limits_{j=1}^{N_a} \sum\limits_{l=1}^{N_a} C_{jli}\beta_{jl} \left(\frac{n_{ij}^{\rm s}}{n_{j}^{\rm d}}+\frac{n_{il}^{\rm s}}{n_{l}^{\rm d}}\right) 
n_{j}^{\rm d} n_{l}^{\rm d} n_{\rm gas} \nonumber \\ 
& - n_{iz}^{\rm s}\sum\limits_{j=1}^{N_{a}}\beta_{jz}  n_j^{\rm d} n_{\rm gas} \Bigg).
\label{eq:ns_corr}
\end{align}

\subsection{Rate coefficients}

The rate coefficients for monomolecular, bimolecular, and termolecular reactions were calculated using the modified Arrhenius law with kinetic parameters $A$, $n$ and $E_{\rm a}$ (the activation barrier), which were defined for each reaction individually (see Sect.~\ref{network}). The adsorption rate  coefficient was calculated as
\begin{equation}
\label{accr}
k^{\rm ad}_{iz} = 4\pi a_z^2 v_i^{\rm th} \alpha_i n^{\rm d}_z 
,\end{equation}
where $v_i^{\rm th}$ is the thermal velocity of the $i$-th molecule and $\alpha_{i}$ is its sticking coefficient. We assume that all hydrocarbons contribute to dust grain growth proportionally to their gas-phase abundances. The most abundant species, such as acetylene (\ce{C2H2}), play the dominant role. We adopted $\alpha_{i}=1$ for all molecules. We accounted for adsorption onto sites occupied by \ce{sC2H2} or \ce{sC4H4} separately, as they are intermediates in CTM (see below). Therefore, the rate coefficient for \ce{C2H2} adsorption was multiplied by $f$, the fraction of sites not occupied by these species. We consider the adsorption process to be a direct chemisorption with no activation barrier.

The desorption rate coefficient is
\begin{equation}
k^{\rm des}_{iz} = \nu_0 \exp \left(-\frac{E_{\rm d}^i}{k_{\rm B}T^{\rm d}_z}\right),
\end{equation}
where $\nu_0$ is a characteristic frequency \citep[see][]{hasegawa92}, $E_{\rm d}^i$ is the desorption energy of the $i$-th molecule, and $T^{\rm d}_z$ is the temperature of the $z$-th dust grain. 

The value of $E_{\rm d}^i$ is a cornerstone in our astrochemical model, so we considered its choice in more detail. Hydrocarbons first adsorb onto SiC grain surfaces. The SiC surface is chemically active for forming strong bonds with hydrocarbons. Calculations presented in \citet{weiferink06, pollmann14} provide values of 1--5~eV depending on the molecular species and the considered reconstruction of SiC surface. Such values strongly support the chemical nature of bonding between the acetylene molecule and the SiC surface. For such a scenario, the desorption process proceeds less effectively, as more energy is required for desorption of adsorbed acetylene, being in a stable configuration.

After the formation of a carbonaceous layer on SiC grains, subsequent adsorption occurs on this carbonaceous surface. While covalent bonding can occur at specific sites, adsorption is not universally chemisorptive. Bond strength critically depends on local surface structure and defect presence (e.g., radicals, zigzag edges). Whereas basal graphene planes exhibit low reactivity, PAH zigzag edges and carbon radicals are highly active sites \citep{radovic05}. \citet{krueger96} argued that adsorption of \ce{C2H2} results in the formation of a sp$^3$ bond, which finally leads to grains having an amorphous structure. Thus, circumstellar environments likely produce irregular carbon surfaces with numerous potential binding sites. 

Determining a single desorption energy for all hydrocarbons and surfaces is challenging due to multiple crucial factors influencing it. In the absence of direct measurements, we can rely only on indirect estimations. Among them, one can mention the work of \citet{alamdari23}, who studied ethylene adsorption on graphenes and gave $E_{\rm d}^{\ce{C2H4}}$ in the wide range 1--5~eV. Based on the results of \citet{gordeev20} who investigated the kinetics of acetylene CTM on carbene active sites, one may conclude that the dissociation of the bond between acetylene and carbene site needs around 2~eV. 

From the works on laboratory formation of interstellar dust analogs, we can conclude that the structure of the dust grains corresponds to hydrogenated amorphous carbon with a mixture of sp, sp$^2$, and sp$^3$ bonds. The strength of these bonds also varies in a wide range, $\sim2-5$~eV~\citep{ROBERTSON2002, SUGIURA2020}, depending on the sp$^2$/sp$^3$ ratio. Additionally, some estimations of binding between aromatic molecules can be borrowed from works on the soot inception. It is believed that the covalent coalescence takes place during the initial soot inception~\citep{GENTILE2020}. Pure van der Waals interaction would yield unstable particles. \citet{frenklach20} identify E-bridge formation (two aliphatic C--C bonds) as the optimal pyrene clustering pathway, with DFT calculations showing a bond strength of $\sim$2~eV. 

Given the uncertainties discussed above, we performed calculations for $E_{\rm d} = \{2,\, 3,\, 4\}$~eV, adopting the intermediate value $E_{\rm d} = 3$~eV as a standard one\footnote{The same value of $E_{\rm d}$ was adopted for all hydrocarbons; therefore, we omitted the index ``$i$''.}. The range 2--4~eV encompasses energies expected for desorption of surface species from SiC or carbonaceous dust grains.

We included two surface reactions of Eley-Rideal type -- s\ce{C2H2}$+$\ce{C2H2}$ \rightarrow $s\ce{C4H4} and   s\ce{C4H4}$+$\ce{C2H2}$ \rightarrow $s\ce{C6H6}. Herewith we adopted that the formed s\ce{C6H6} desorbs without a barrier immediately after cyclization, which seems to be plausible based on our DFT calculations of acetylene CTM reaction on the SiC surface (Fedotova et al., in prep.). Should we consider the CTM process in more detail, then the formation of benzene from three acetylene molecules can proceed via at least two distinct reaction channels. The first involves the linear growth of an acetylene chain to form s\ce{C6H6}, followed by cyclization. The second pathway proceeds through the initial formation and surface stabilization of s\ce{C4H4} ring, which a subsequent acetylene molecule then opens and extends to form cyclic s\ce{C6H6}. In the present model, we do not computationally distinguish these two channels; their kinetics are treated as identical. A detailed discussion of the differences will be provided in a forthcoming paper by Fedotova et al. (in prep.). It should be mentioned that we assume that after forming the carbonaceous layer on SiC dust grains surface reactions proceed further with the same formalism, meaning that carbon dangling bonds act as active sites instead of Si in this case.

In the Eley-Rideal mechanism, a reaction occurs directly between a gas-phase species and an adsorbed species on a surface. The rate coefficient for such a reaction is analogous to the adsorption rate coefficient, but must be computed for a single surface site. This is achieved by dividing the adsorption rate coefficient by the total number of sites available on a grain of a specific size, $n^{\rm d}_z n_{{\rm sites},z}$. Thus, the rate coefficient for the Eley-Rideal reaction between the $j$-th adsorbed molecule and a gas-phase acetylene molecule is given by
\begin{eqnarray}
\label{ksr}
k_{ijz} & = & 4 \pi a_{z}^2 v_{\ce{C2H2}}^{\rm th} \alpha_{\ce{C2H2}}n^{\rm d}_z \frac{1}{n^{\rm d}_z n_{{\rm sites},z} } \nonumber \\
\text{\;or} && \nonumber \\
 k_{ijz} & =&  4 \pi a_{z}^2 v_{\ce{C2H2}}^{\rm th}  \alpha_{\ce{C2H2}}\frac{1}{n_{{\rm sites},z} } 
.\end{eqnarray} 

As was mentioned, apart from the acetylene CTM on the SiC surface, we also considered this reaction on zigzag edges of PAHs. According to \cite{gordeev20}, PAHs with carbene centers can be efficient catalysts for CTM of acetylene. The approximate value of $E_{\rm a}$ is $\approx 0.5$~eV of the first and subsequent acetylene attachments, and its free energy is $\Delta G \approx -1$~eV. Cyclization proceeds with $E_{\rm a}\approx 0.2$~eV and $\Delta G \approx -2$~eV. After cyclization, the formed benzene molecule can be desorbed either directly or after another acetylene becomes associated with the same site. For small PAHs, the first way is more attractive because of lower values of $E_{\rm a}$, which are $\sim 0.8-0.9$~eV (C$_6$, C$_{14}$) against $\approx1.5$~eV for C$_{37}$. In its turn, the second way is more favourable for large PAHs as it requires energy around 1.2~eV almost independently of the PAH size.

We adapted the results of the work of \cite{gordeev20} in the following way: 1) only the largest PAH in our network, pyrene (radical), acts as a catalyst; 2) we added four reactions for pyrene: C$_{16}$H$_9^{\bullet} +\ce{C2H2} \rightarrow $ C$_{16}$H$_9$-\ce{C2H2}, C$_{16}$H$_9$-\ce{C2H2} +\ce{C2H2} $\rightarrow $ C$_{16}$H$_9$-\ce{C4H4}, C$_{16}$H$_9$-\ce{C4H4} +\ce{C2H2} $\rightarrow $ C$_{16}$H$_9$-\ce{C6H6}, and C$_{16}$H$_9$-\ce{C6H6}$\rightarrow $ C$_{16}$H$_9^{\bullet}$ +\ce{C6H6}; 3) $E_{\rm a}$ for acetylene addition was adopted to be 0.5~eV; 4) the energy of \ce{C2H2}- and \ce{C4H4}-loss, $E_{\rm d}^{\rm PAH}$, is 3~eV; 5) the energy of \ce{C6H6}-loss is 1~eV. The expression for rate coefficients for the reactions of the formation of the intermediates C$_{16}$H$_{10}$-C$_2$H$_2$, C$_{16}$H$_{10}$-C$_4$H$_4$, and C$_{16}$H$_{10}$-C$_6$H$_6$ is analogous to Eq.~\ref{ksr}, except that we multiplied the adsorption rate by a factor of $\exp(-E_{\rm a}/(k_{\rm B}T^{\rm PAH}))$, accounting for the activation barrier of the attachment. To find the rate of the reaction, the rate coefficient was multiplied by the number densities of the corresponding PAH and gaseous acetylene.  

\subsection{The reaction network}
\label{network}

There are several networks presented in the literature that are used to model the chemistry of AGB stellar envelopes. \citet{frenklach89} in their pioneering work presented the network of reactions, which was mainly borrowed from combustion studies. This network and the calculations performed on its base become fundamental in modeling of aromatics in stellar envelopes. Later, \cite{cherchneff92} corrected the network, supplementing it with new studied reactions and neglecting the reactions that have little importance. Many further studies have relied on the results obtained with these networks~\citep[e.g.][]{allain97, pascoli00, cherchneff12}, so the significance of the networks is undeniable. However, kinetic parameters for many reactions in these networks were poorly estimated or even extrapolated; namely, for the reactions of PAH growth via the HACA mechanism. 

The most recent network of reactions was presented in \cite{cherchneff12}, but as far as aromatics are concerned it includes only reactions leading to formation of benzene. Thus, we took this network as a base for a new network and supplemented it with a number of reactions of PAH growth up to pyrene for which kinetic parameters are either available or could be estimated. We also revised the Cherchneff's network and corrected kinetic parameters for some reactions if more accurate data are present in literature.  

In our network, the first aromatic ring can be formed via following routes in the gas phase:
\begin{itemize}
    \item [1.] Recombination of propargyl radical resulting in the formation of benzene: 
    
    \ce{C3H3}$^{\bullet} +$ \ce{C3H3}$^{\bullet} \rightleftharpoons $ a-\ce{C6H6}\footnote{The prefix ``a-'' means ``aromatic'' in our notation system to distinguish aromatic and non-aromatic molecules.}, 
    
    for which we used the kinetic parameters given in \cite{zhao21}, and in the formation of a phenyl radical:
    
    \ce{C3H3}$^{\bullet} +$ \ce{C3H3}$^{\bullet} \rightleftharpoons $ a-\ce{C6H5}$^{\bullet}$+H, 
    
    with kinetic parameters from \cite{cherchneff12}. The phenyl radical further forms benzene via collisions with H, H$_2$, or H-bearing species.   
    
    \item [2.] Isomerization of linear \ce{C6H6} (2-ethynyl-1,3-butadiene) to benzene directly: 
    
    \ce{C6H6} $\rightleftharpoons $ a-\ce{C6H6},
    
    or to a phenyl radical:
    
    \ce{C6H6} $\rightleftharpoons $ a-\ce{C6H5}$^{\bullet}$+H.
    
    \item [3.] The reactions of 1,3-butadien-1-yl (denoted as n-\ce{C4H5}) and 1,3-butadien-2-yl (denoted as i-\ce{C4H5}) with acetylene: 
    
    n-\ce{C4H5} $+$ \ce{C2H2} $\rightleftharpoons $ a-\ce{C6H6} $+$ H \citep{wang97} and 
    
    i-\ce{C4H5} $+$ \ce{C2H2} $\rightleftharpoons $ a-\ce{C6H6} $+$ H \citep{huang18}.
    \item[4.] The reaction between 1,3-hexadien (denoted as l-\ce{C6H6}) and H, which gives either a cycled \ce{C6H7} with additional hydrogen, which can dissociate later, 
    
    l-\ce{C6H6}+H $\rightleftharpoons $ a-\ce{C6H7} \citep{wang97}, 
    
    or linear 1,3,5-hexatrien-1-yl, which can later cyclysize: 
    
    l-\ce{C6H6}+H $\rightleftharpoons $ \ce{C6H7} \citep{wang97}, 
    
    \ce{C6H7} $\rightleftharpoons $ a-\ce{C6H7} \citep{duran88}.

    \item[5.] Isomerization of fulvene to benzene: 

    a-\ce{C5H4}-\ce{CH2}\footnote{``-R'' means a functional group.} $\rightleftharpoons $ a-\ce{C6H6} \citep{zhao21}.

    In its turn, fulvene can be formed either from phenyl radical \citep{zhao21} or in the reaction between i-\ce{C4H5} and \ce{C2H2}~\citep{huang18}. 

    \item[6.] The reaction between 1-buten-3-yn-lyl (denoted as n-\ce{C4H3}) and acetylene results in formation of a phenyl radical,

    n-\ce{C4H3} $+$ \ce{C2H2} $\rightleftharpoons $ a-\ce{C6H5}$^{\bullet}$ \citep{wang97},

    or benzyne,

    n-\ce{C4H3} $+$ \ce{C2H2} $\rightleftharpoons $ a-\ce{C6H4} +H \citep{wang97},

    which further may form benzene in reactions with H-bearing species as well as phenyl radical.
    
\end{itemize}

We consider two molecules, consisting of two aromatic rings -- biphenyl and naphthalene. Biphenyl forms out of benzene and phenyl molecules and is then converted to phenanthrene. Naphthalene is formed via the HACA mechanism through subsequent steps of hydrogen abstraction and acetylene addition. On the way, two intermediates can appear -- a-\ce{C6H5}-\ce{C2H2} and a-\ce{C6H4}-\ce{C2H3}. The kinetic parameters were taken from \citet{mebel17}. The path to naphthalene synthesis is a dead end in our network, as it stops after the formation of acenaphthylene. This limit of the HACA mechanism was found in a number of works \citep{kislov05, parker15}, although it has already been shown that naphthalene can grow further via the phenyl-addition -- dehydrocyclization mechanism, which implies the addition of phenyl radical and formation of triphenylene~\citep{zhao19}. There are also other ways to form naphthalene from benzene~\citep{mebel17}; for example the hydrogen abstraction-vinylacetylene addition mechanism, in which the growth proceeds through addition of vinylacetylene instead of acetylene~\citep{parker12, kaiser15}, but this mechanism is probably operative more efficiently at low temperatures, so we keep it outside the scope of the work.

One more benzene derivative -- toluene, forming in the reaction a-\ce{C6H6} + \ce{CH3} -- participates in formation of anthracene and phenanthrene according to \citet{sinha16, kaiser22}. Two toluene radicals then recombine and form H-rich complexes with either phenanthrene or anthracene geometry, which subsequently lose their hydrogen atoms and transform into phenanthrene or anthracene. Finally, pyrene is formed from phenanthrene via the HACA mechanism, which has been studied in \citet{frenklach19}.

Obviously, this reaction network is far from being complete up to pyrene and molecules of comparable size. In addition to the above-mentioned omitted pathways, we have not considered the growth reactions of PAHs with five-membered rings, such as indene, considered in \citet{mebel17}. Besides the incompleteness, there are several inaccuracies that can be improved in the future. One of them is using a kind of a lumping method widely used in combustion theories \citep[e.g.,][]{frenklach91}. We applied it when two isomers (e.g., \ce{C4H3}) potentially had different rate coefficients for the same reactions or even participated in different reactions, but the kinetic parameters for either isomer were not available. In these cases we used the same rate coefficients for both isomers if they were reactants and a half of the rate coefficient value if they were products.

A general uncertainty for most of the reactions is related to the difference of the atmosphere composition, at which DFT-calculations were performed -- it is N$_2$, characteristic of the Earth, versus H$_2$ for stellar envelopes. This issue is discussed separately in Appendix~\ref{n2}. 

Another source of uncertainty stems from the discrepancy between the low pressures typical of stellar envelopes (below $\sim10^{-3}$~atm) and the pressures at which the kinetic rates are typically reported. Although we selected rate coefficients corresponding to the lowest available pressure values, these rarely extend below 0.1~atm. To improve the accuracy of the reaction network, a comprehensive recalculation of rate coefficients at lower pressures would be necessary -- a task that remains computationally demanding.

An additional uncertainty arises from the treatment of reverse reaction rates. When these rates are not explicitly provided, we estimated them using the corresponding forward reaction rates and equilibrium constants derived from thermochemical data given in \citet{wang97} or GRI-Mech\footnote{\url{http://combustion.berkeley.edu/gri-mech/version30/text30.html}}. While this approach prevents the artificial overaccumulation of certain reaction products, it introduces greater uncertainty in our results since our physical conditions are far from thermodynamic equilibrium.

Additionally, we have excluded photoreactions from our model, as we assume that UV radiation plays a negligible role in hydrocarbon chemistry within the studied envelope regions. This assumption is justified by two factors: (1) external UV radiation is absorbed efficiently in the outer regions to be significant, and (2) stellar emission of IRC+10216 in this spectral range is insufficiently intense. While comprehensive photochemical networks exist \citep[notably the UMIST database for stellar envelopes;][]{millar00, umist13, millar16}, these are primarily applicable to outer envelope layers where interstellar UV photons dominate the chemistry -- a regime outside our current scope of investigation.

\subsection{Treatment of dust evolution}

The presence and properties of dust grains are fundamental to this study. As was previously noted, our model does not include homogeneous SiC nucleation. We assume that SiC seed particles first form in the stellar vicinity and present in the envelope at radial distances $R \geq 2\,R_{\star}$, where conditions permit both their survival and subsequent growth. We adopted a size distribution for SiC seed particles following the law $n^{\rm d}(a) \propto a^{-3.5}$ \citep[the MRN distribution,][]{mrn}. While the MRN distribution was originally proposed for the ISM dust, we employed it here as it represents a well-established standard that has been applied to stellar envelope modeling in previous works \citep[see, e.g.,][]{speck09}. Test calculations performed under the assumption of an equal grain distribution showed that while the adopted dust size distribution influences hydrocarbon abundances, the general trends and conclusions remain unchanged. These results are provided in Appendix~\ref{app: abund_difmod}.

We divided grains into  $N_{a}=15$ size bins spanning the range from 0.005 to 0.25~$\mu$m. Initially, only bins corresponding to radii $a\leq0.1~\mu$m were populated, with bins of larger sizes being filled gradually over time due to adsorption and coagulation. Our choice of 0.1~$\mu$m as the upper limit for initial seed particles is motivated by several factors. Firstly, theoretical models of dust growth suggest limited efficiency for particles in excess of this size~\citep{cadwell94, krueger97}. Secondly, this value represents a standard choice for grains in monodisperse models or for an upper size limit in models with multiple grain populations used for SED calculations of AGB stars~\citep{lorenz94, suh00}. Finally, observations of IRC$+$10216 indicate that dust sizes are mostly below 0.1~$\mu$m~\citep{martin87}. Unlike estimates of the circumstellar dust size, meteoritic studies reveal larger SiC grains, which are believed to have circumstellar origin. Typically they are 0.3--0.7~$\mu$m, but some of them reach several microns~\citep{daulton03}. However, these measurements likely reflect survival during traveling across the ISM and evolution in the Solar System rather than formation in stellar envelopes.

The dust-to-gas mass ratio in stellar envelopes is estimated to have a range from $\sim10^{-3}$ to $\sim5\cdot10^{-3}$~\citep{groenewegen97, mattsson10, cherchneff12,  hofner18}. Within this dust amount, SiC grains typically constitute $1-10$\% by mass~\citep{speck09, gomezllanos18}, yielding an $m_{\rm SiC}/m_{\rm gas}$ ratio between $10^{-5}$ and $10^{-4}$. In our model, we adopt $m_{\rm SiC}/m_{\rm gas}=10^{-4}$ as a representative value. Assuming a SiC bulk density of $\rho_{\rm SiC}=2.5$~g~cm$^{-3}$ and the MRN size distribution, the ratio $m_{\rm SiC}/m_{\rm gas}$ corresponds to an initial seed number density of $5\cdot10^{-12}$ relative to $n_{{\rm gas}}$. This value agrees well with grain number densities predicted by theoretical models of dust growth~\citep{cadwell94, krueger97, gail_book}. Note that we assume that homogeneous SiC grain growth terminates once carbonaceous species begin to adsorb onto their surfaces.

 The dust temperature was calculated from the equation of thermal balance, assuming that dust is heated by the stellar radiation and by thermal collisions with gas particles and is cooled through infrared emission.  Details on the calculations are given in Appendix~\ref{dust_temp}. In order to find the dust temperature, absorption and scattering efficiencies ($Q_{\rm abs}$ and $Q_{\rm sca}$) for grains materials are required. Those were calculated from known refractive indices  with the SIGMA program\footnote{\url{https://github.com/charlenelefevre/SIGMA}}~\citep{lefevre19_sigma, min05, min16}. Refractive indices are taken from \citet{laor93} for SiC particles and from \citet{zubko96} for amorphous carbon. We distinguish purely SiC grains and grains with a SiC core and carbonaceous mantle.  A proportion in which each material contributes to the cross section depends on the mass of adsorbed molecules relative to the core mass and changes at each calculation step. The high temperature of dust grains and correspondingly high desorption rate in the beginning of the pulsation phase make abundances of surface species very small, but they increase again by the end of the phase.

Thorough modeling of dust evolution requires a more sophisticated model. In particular, our model includes dust coagulation but we do not consider the reverse process, i.e., fragmentation due to uncertainties in parameters of this process. To test the possible significance of fragmentation, we performed additional calculations assuming that dust size distribution reverses back to the initial one at the beginning of each shock wave. It was found that our conclusions do not differ substantially in this case, which we show in Appendix~\ref{app: abund_difmod}.

In addition to fragmentation, we neglected structural transformations of the material that may occur at elevated temperatures during shock wave passage. While amorphous carbonaceous particles undergo dehydrogenation and graphitization at dust temperatures of $T^{\rm d} \gtrsim 2000$~K~\citep{jones22}, followed by sublimation of carbon atoms or molecules from the graphitized surface (a process distinct from our adopted desorption treatment), we excluded these effects from our model. This simplification is justified because: 1) the duration of $T^{\rm d} > 2000$~K conditions is extremely brief ($\sim$1~s or less); 2) such high temperatures occur only in the immediate stellar vicinity ($R \approx 2.2\;R_{\star}$); 3) graphitization significantly affects only the smallest dust particles and only partially. 

\section{Results}

\subsection{Abundances of aromatic molecules}

We performed calculations of chemical abundances in the envelope with conditions corresponding to IRC+10216 within the region between 1.2 and 5~$R_{\star}$. Models with surface reactions (i.e., with CTM on the dust surface) and without them are considered. 

\begin{figure*}
\centering
\includegraphics[width=0.45\textwidth]{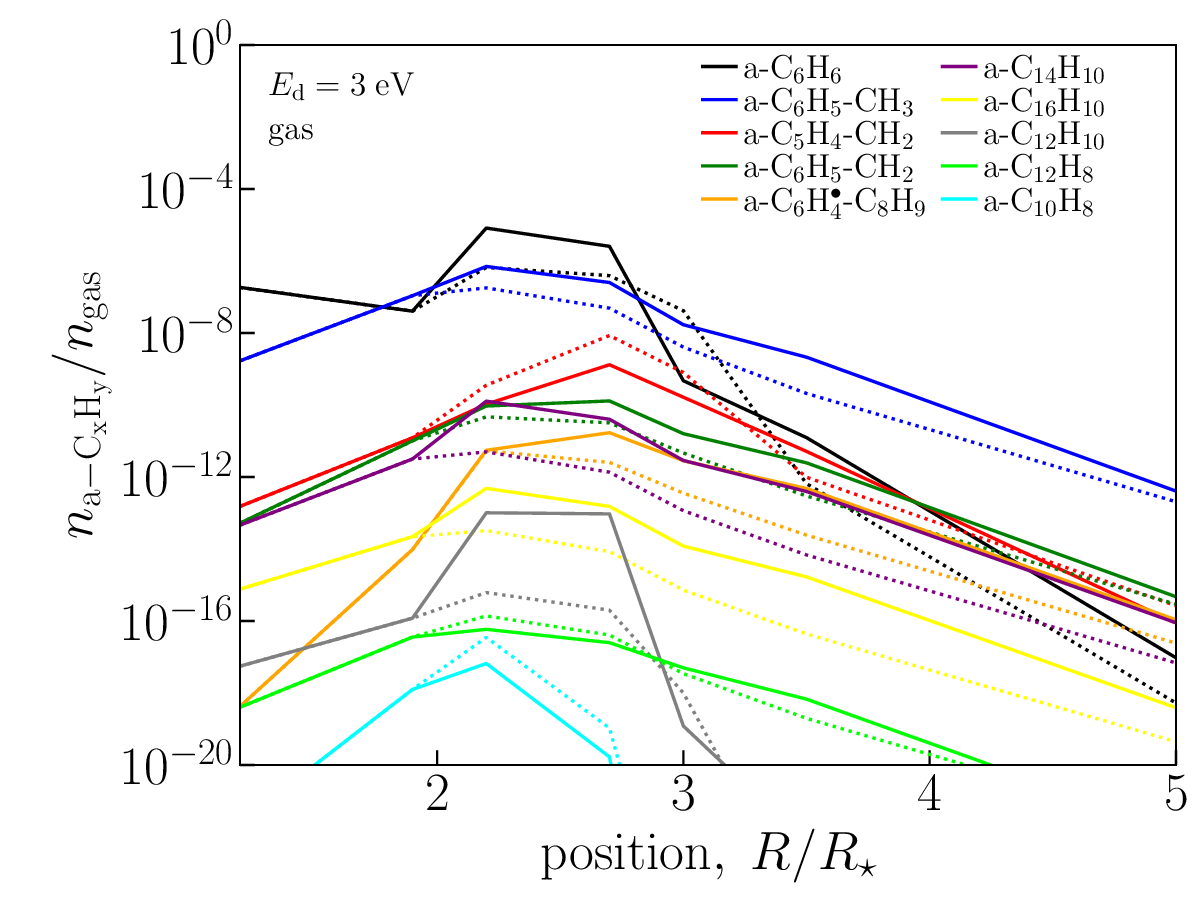}
\includegraphics[width=0.45\textwidth]{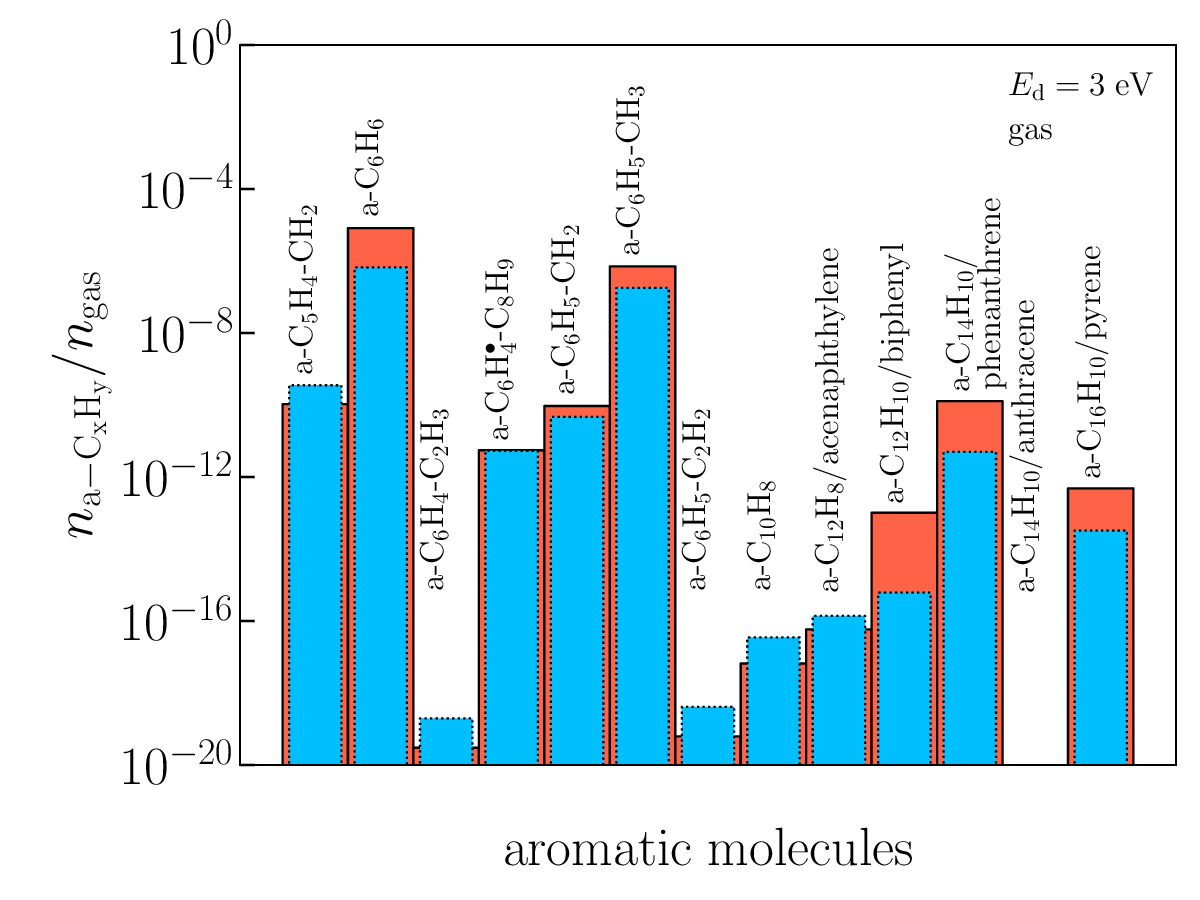}
\caption{Left: Gas-phase abundances of aromatic molecules versus distance from the star calculated with CTM (solid lines) and without it (dotted lines). Right: Gas-phase abundances of aromatic molecules at $R=2.7\;R_{\star}$. The red color indicates the model with CTM. The blue color shows the model without it.}
\label{abund_gas}
\end{figure*}

In Fig.~\ref{abund_gas}~(left) we demonstrate gas-phase abundances of aromatic molecules versus distance from the star. The abundances correspond to the end of the adiabatic phase at each position. Both models with CTM (solid lines) and without it (dotted lines) are presented. It is seen that in both cases abundances behave in a similar way. Their maxima fall around 2--3~$R_{\star}$ which is consistent with the conclusion of \citet{cherchneff12} that benzene is most efficiently formed at this distance. Abundances mostly are higher in the model with CTM, although some molecules (such as fulvene, naphthalene, and even benzene\footnote{Note that in Fig.~\ref{abund_gas} and further the designation `\mbox{a-\ce{C6H6}}' includes not only benzene itself, but also phenyl radical (a-\ce{C6H5}$^{\bullet}$) and benzyne (\mbox{a-\ce{C6H4}}).} at $R = 3\;R_{\star}$) can be more abundant in the purely gas-phase model. At $R\gtrsim 3\;R_{\star}$ abundances gradually decrease for two reasons: 1) the rate of consumption of molecules in reactions is higher than their formation rate; 2) the molecule adsorb onto dust surfaces, and their desorption is ineffective at decreasing temperature. Abundances of benzene, fulvene,  naphthalene, and biphenyl fall steeper than abundances of other molecules, because the difference between the rates of their consumption and formation is  bigger.

At $R= 3\;R_{\star}$, the abundance of benzene predicted by the model without CTM is higher than the value predicted by the model with CTM. According to \citet{cherchneff12}, it is at this point that the abundance of benzene formed in gas-phase reactions reaches its maximum. As our network of reactions up to benzene is very close to Cherchneff's network, this statement is valid for our work in the model without CTM. However, if CTM is included, then part of acetylene had been already consumed in it or locked in \ce{sC4H4}. Thus, the amount of available acetylene turns out to be lower. At larger distances, when efficiency of gas-phase reactions decreases, the abundance of benzene predicted by the model with CTM exceeds the value predicted by the model without CTM.

To demonstrate difference between models more clearly, we present the abundances of aromatic molecules at $R= 2.7\;R_{\star}$ as a histogram in Fig.~\ref{abund_gas}~(right). The most noticeable difference (around an order of magnitude) is seen for benzene, biphenyl, phenanthrene, and pyrene. Thus, the addition of surface CTM accelerates the formation of these molecules. In its turn, naphthalene and its precursors (\mbox{a-\ce{C6H5}-\ce{C2H2}}, \mbox{a-\ce{C6H5}-\ce{C2H3}}) are slightly more abundant in the model without CTM because in this case acetylene is not consumed in CTM and consequently is available for other reactions. 

\begin{figure*}
\centering
\includegraphics[width=0.45\textwidth]{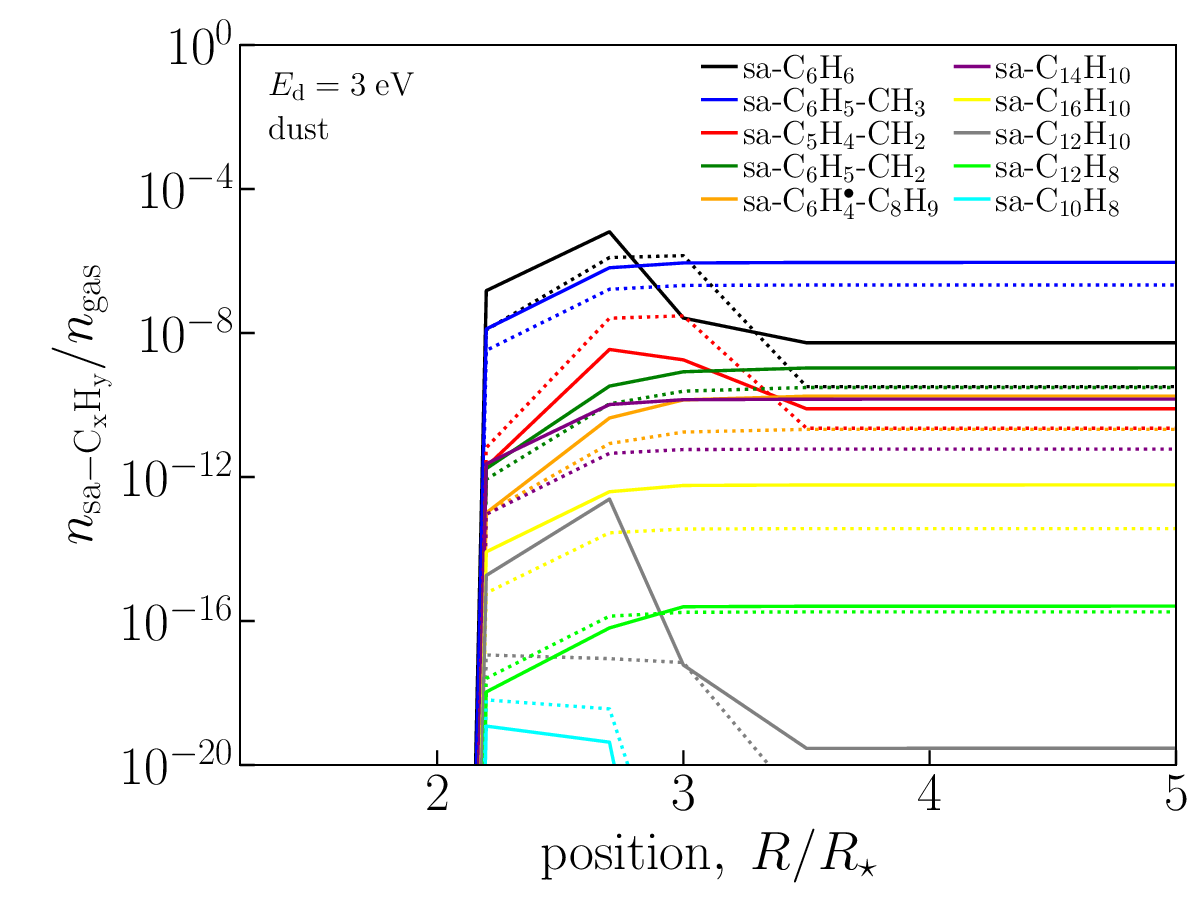}
\includegraphics[width=0.45\textwidth]{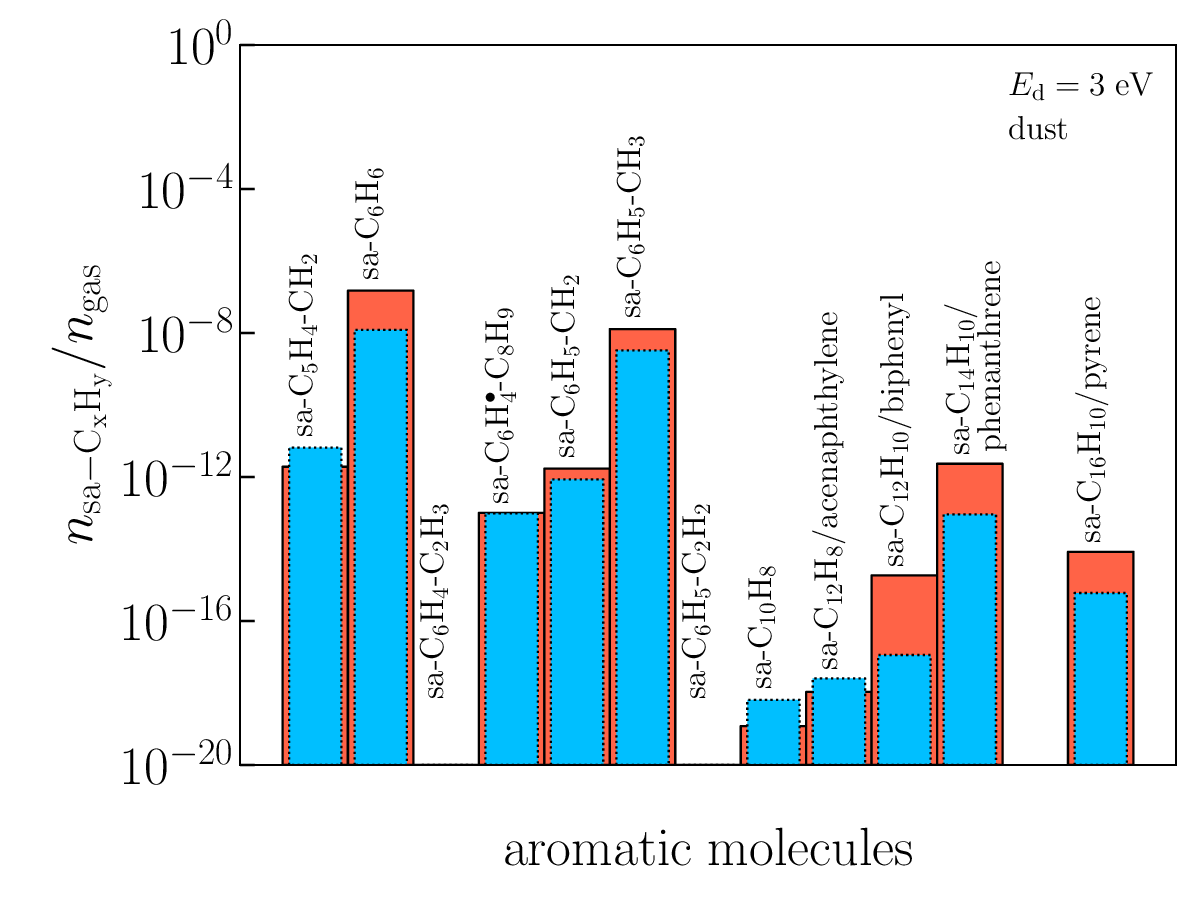}
\caption{Same as in Fig.~\ref{abund_gas} but abundances of aromatic molecules on surfaces of dust grains are presented.}
\label{abund_dust}
\end{figure*}

In Fig.~\ref{abund_dust} we demonstrate the abundance of aromatic molecules on surfaces of dust grains. We remind the reader that it is assumed that there is no dust at $R <2\;R_{\star}$; therefore, surface abundances of molecules are zero. From the left panel of Fig.~\ref{abund_dust}, it is seen that the abundances of some molecules (toluene (\mbox{a-\ce{C6H5}-\ce{CH3}}), phenanthrene, pyrene) achieve their maxima at $R\approx 3\;R_{\star}$ and remain constant up to $R=5\;R_{\star}$. Abundances of benzene, fulvene, naphthalene, and biphenyl reach the maxima at $R=2.7\;R_{\star}$, which approximately coincides with the position of their maxima in the gas phase. Further from the star their abundances on the dust surface decrease, which is consistent with that their gas-phase abundances fall steeper than for other molecules at these distances. The histogram on the right panel of Fig.~\ref{abund_dust} is similar to that in Fig.~\ref{abund_gas}, but abundances of molecules on the surface are lower than in the gas phase at $R=2.7\;R_{\star}$. Note that a detailed analysis of the dust size distribution evolution is beyond the scope of this study; however, a related figure is provided in Appendix~\ref{app: dust_distr}.

\subsection{Total abundance of aromatics and its sensitivity to the desorption energy}

\begin{figure*}
\centering
\includegraphics[width=0.45\textwidth]{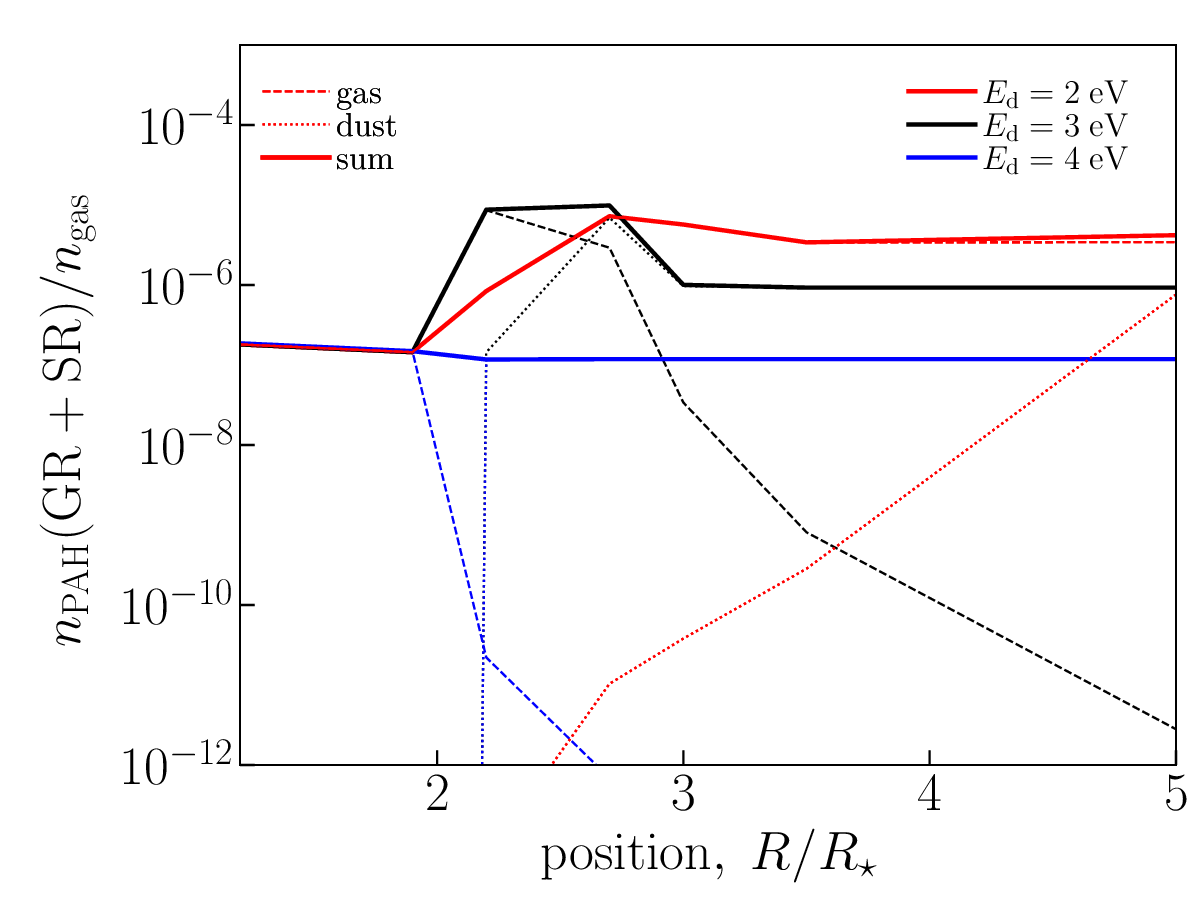}
\includegraphics[width=0.45\textwidth]{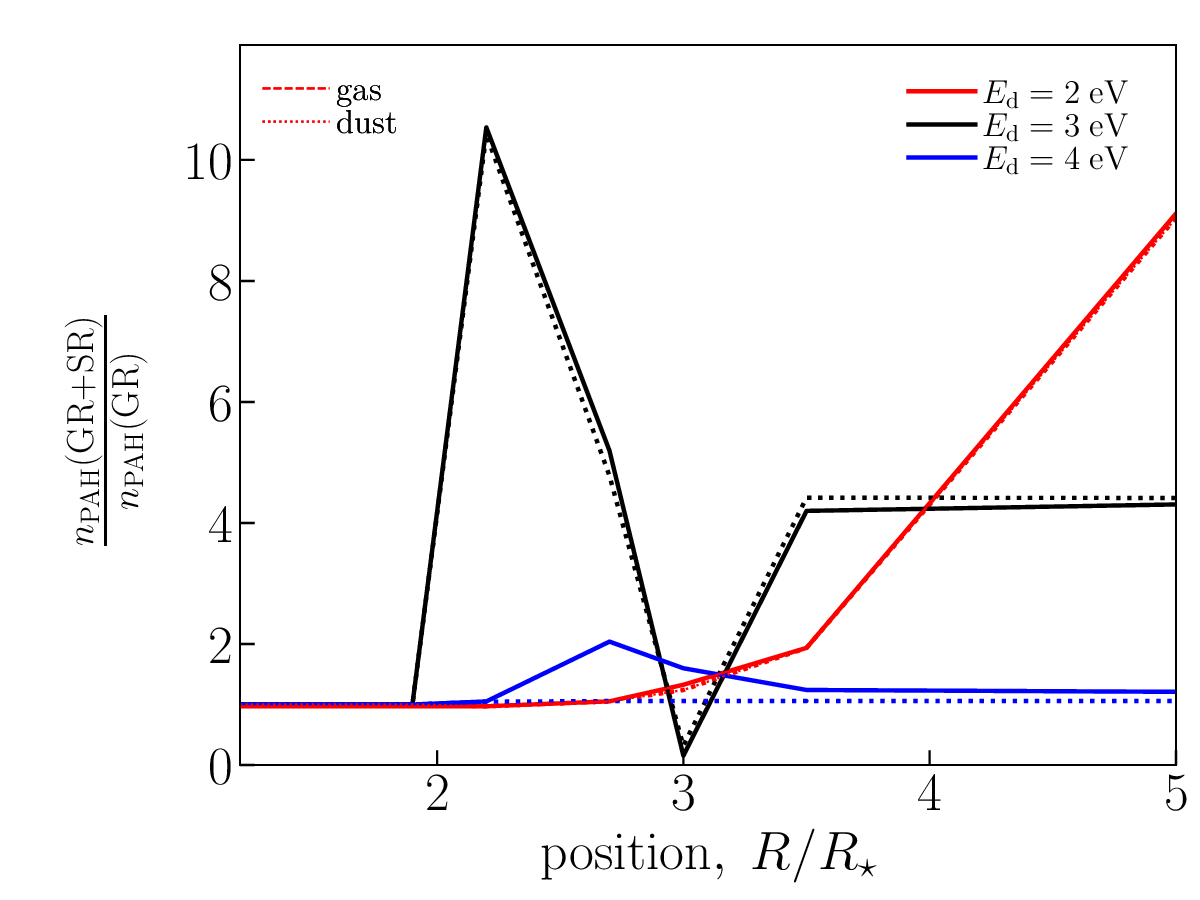}
\caption{Left: Total abundance of all aromatic molecules obtained in the model with CTM. Gas-phase abundances are shown by dashed lines, solid-phase abundances are shown by dotted lines, and the sum abundance is shown by solid lines. Red, black, and blue colors illustrate the models with $E_{\rm d}=2$, 3 and 4~eV, respectively. Right: Ratio between the total abundance of all aromatic molecules obtained in the models with the surface acetylene CTM and without it. The designation ``GR+SR'' means the model with CTM, while the designation ``GR'' indicates the model without it. The ratio of gas-phase abundances is illustrated by solid lines, while the ratio of solid-phase abundances is shown by dotted lines. The colors correspond to the same models as in the left panel.}
\label{sumabund}
\end{figure*}

In order to analyze the results of our modeling and to estimate the contribution of the surface CTM, we calculated the total abundance of aromatics ($n_{\rm PAH}$) in gas and on dust at each position. The abundances on dust are presented to show the total abundance of formed aromatics independently of their current phase as the state can change when some influencing factors (such as shock waves and UV radiation) appear. We illustrate the variation of the total abundance with the distance at different  $E_{\rm d}$ in Fig.~\ref{sumabund}~(on the left). Additionally, we give the values in Table~\ref{table:summary}. As in previous figures, the abundances correspond to the end of the adiabatic phase at each position. 

\citet{cherchneff12} performed simple estimations of the total PAH and carbonaceous dust mass assuming that all benzene convert to coronene molecules and that all these coronene molecules are involved in dust grain growth. She did not perform detailed calculations of all thermal pulses that parcels may undergo. Instead she multiplied the maximum values of PAH and dust masses by an approximate number of thermal pulses, implying that each one leads to formation of the same amount of benzene and, consequently, coronene and more complex carbonaceous dust. The obtained dust-to-gas ratio was found to be in the range $1.2\cdot10^{-3}-5.8\cdot10^{-3}$, which is consistent with observational values and theoretical predictions of $\approx 10^{-3}-5\cdot10^{-3}$~(\cite{groenewegen97}, etc.). We did not repeat the procedure of \citet{cherchneff12} because we do not aim to reproduce the observed dust mass. Instead, we focus mainly on the PAH formation process and its variation with distance. Therefore, we discuss the calculated abundances without extrapolation of these values to other pulsations. Thus, it should be kept in mind that our total abundances can be underestimated due to the lower number of pulses included in the model (it is 7) in comparison with the real number of pulses \citep[from 9 to 43 in][]{cherchneff12}.

In Fig.~\ref{sumabund}, $n_{\rm PAH}$ at $E_{\rm d}=3$~eV reaches its maximum at $R\sim 2-3\; R_{\star}$. This region is predicted to be the most favorable for dust formation~\citep{krueger97, gail_book}. At $R=2.2 \;R_{\star}$, aromatic molecules are mainly in gas, but they adsorb intensively on the dust surface when moving away from the star. At $R=5 \;R_{\star}$, the gas phase $n_{\rm PAH}$ is only $2.80\cdot10^{-12}$ relative to $n_{\rm gas}$. Certainly, when the parcel undergoes another shock wave, part of aromatics is ejected into gas again. Nevertheless, under the conditions corresponding to $R=5 \;R_{\star}$ and at $E_{\rm d}=3$~eV gas components tend to be solidified. Overall, $n_{\rm PAH}$ reaches $9.86\cdot10^{-6}$ at $2.7\; R_{\star}$, which exceeds the maximum abundance of benzene predicted by \citet{cherchneff12}. \citet{cherchneff12}'s value of $n_{\ce{C6H6}}$ is 1.31$\cdot10^{-6}$ relative to $n_{{\rm H}_2}$ ($n_{{\rm H}_2}$ is generally very close to $n_{\rm gas}$), which is closer to the maximum $n_{\rm PAH}$ predicted by the model without CTM.  

It is not possible to compare the predicted PAH abundance with observations as PAHs in stellar envelopes are not excited and consequently are not observed. However, we may compare the predictions for $n_{\ce{C2H2}}$ with observations, although such a comparison is also not straightforward. The observations provide the summed abundances over the phase when shock waves pass through the parcel and most of the surface components evaporate into gas and the stage when gas components adsorb back on the surface. As $n_{\ce{C2H2}}$ at the beginning and at the end of penetration of a shock wave into a parcel are different, we suppose that it is the mean abundance predicted by the model that should be compared with observations. We provide both abundances of \ce{C2H2} in Table~\ref{table:summary}. 

\cite{fonfria08} reported acetylene abundances of $8\cdot10^{-6}$ (relative to H$_2$) in the inner envelope and $8\cdot10^{-5}$ in outer regions. In the region of efficient PAH and dust formation ($R\sim 2-3\; R_{\star}$), the model with $E_{\rm d}=3$~eV and CTM predicts $n_{\ce{C2H2}}$ in the range $2.92\cdot10^{-7}-10^{-4}$ at the beginning of a shock wave and $2.93\cdot10^{-9}-1.24\cdot10^{-5}$ in its end. The range of values is broad; nevertheless, one may conclude that they satisfy the observations. However, at larger distances $n_{\ce{C2H2}}$ becomes less than $10^{-10}$, which is quite low. If CTM is not included, predicted $n_{\ce{C2H2}}$ is higher, but at large distances $n_{\ce{C2H2}}$ is also low (up to $\approx 10^{-9}$). We suppose that such a high acetylene depletion would be corrected if an activation barrier for chemisorption on surfaces were to be added but we leave investigation of this issue for future work. 

Let us consider the model with $E_{\rm d}=2$~eV. Such a low $E_{\rm d}$ extends the region of efficient PAH formation and dust growth from $2-3\; R_{\star}$ to $5\; R_{\star}$. Thus, the clearly defined PAH and dust formation region is absent. Most theoretical models predict  that dust formation region centers at $2-3\; R_{\star}$ and ends to 5\;$R_{\star}$~\citep{krueger97, gail_book}. Our model with $E_{\rm d}=2$~eV may be in slight contradiction with this statement. Nevertheless, these results represent a model with the lower-bound estimate for $E_{\rm d}$ that remains physically plausible. In this case, maximum $n_{\rm PAH}$ equals to $7.24\cdot10^{-6}$, which is somewhat lower than the maximum value predicted by the model with $E_{\rm d}=3$~eV. However, at $5\; R_{\star}$ $n_{\rm PAH}$ is $4.19\cdot10^{-6}$, which is $\sim5$ times higher than $n_{\rm PAH}$ in the case of $E_{\rm d}=3$~eV. Also, unlike the model with $E_{\rm d}=3$~eV, most molecules are in the gas phase at $5\; R_{\star}$. 

The abundance of acetylene predicted by the model with $E_{\rm d}=2$~eV is quite high -- $\sim 7\cdot10^{-5}-10^{-4}$, i.e., it exceeds the observational estimates given by \citet{fonfria08}. Such a high abundance is consistent with the values of \cite{cherchneff12}, who did not consider participation acetylene in dust growth. But dust growth is included in our model, and acetylene is believed to be one of the main building block in this process. Thus, the model with $E_{\rm d}=2$~eV does not satisfy the necessary rates of dust growth although it provides a reasonable gas-phase amount of PAHs.

The model with $E_{\rm d}=4$~eV represents an upper-bound estimate for the desorption energy. The results obtained with this model are outstanding in comparison with the models considered above. As soon as dust grains appear in the calculations, i.e., at $2.2\; R_{\star}$, aromatics adsorb on their surface. Consequently, the abundance of molecules in gas (aromatic and non-aromatic hydrocarbons) drops, and all aromatics further present only on the surface. The lack of acetylene and other hydrocarbons in the gas phase results in ceasing of PAH formation. CTM also does not work for the same reason. Therefore, there is no PAH formation and dust growth as most hydrocarbons formed in the locations closest to the star are depleted on SiC grains.

Thus, the high $E_{\rm d}=4$~eV case appears to be implausible. Moreover, the results obtained with this model are inconsistent with observed acetylene abundance: $n_{\ce{C2H2}}$ is $\sim10^{-10}$ and lower at $R\gtrsim 3\;R_{\star}$. 

Summarizing the results presented in Fig.~\ref{sumabund}, we can argue that the total abundance predicted by different models varies significantly with the adopted value of $E_{\rm d}$, establishing it as a critical parameter governing the CTM efficiency. While the $E_{\rm d}=4$~eV case provides an upper bound with extreme results, the models with $E_{\rm d}=2$ and 3~eV model better describe the actual hydrocarbon chemistry evolution. However, we suppose that a realistic compromise falls between these two models. 

The uncertainty in observational estimates of the amount of PAHs formed during the AGB phase persists during the post-AGB phase, where the increasing UV radiation from the emerging white dwarf excites PAHs and other small carbonaceous species. This excitation produces diverse mid-infrared spectra, which can be broadly classified into several types~\citep{sloan14}, though many objects defy these categories. Crucially, the ``classical'' PAH emission spectra ubiquitous in the ISM~\citep{peeters02} are almost not observed in these objects. A possible explanation for this absence is that the newly formed PAHs are not isolated in the gas phase. Instead, they are likely bound to the surfaces of dust grains or agglomerated with each other. They also can be substitutes of grains with mixed and irregular structure such as ``aromers''~\citep{homann98} or aromatic-aliphatic linked hydrocarbons (see \citealt{danna09}). This state would prevent PAHs from producing their characteristic emission bands. Therefore, accurately modeling PAH evolution requires one to explicitly account for their interaction with solid surfaces.

Based on our results, we conclude that it is crucial to model the chemistry of hydrocarbons and the dust growth consistently because they are tightly related to each other. Consumption of gaseous hydrocarbons in the dust growth process reduces their availability for PAH formation and growth. While in this work all hydrocarbons are consumed equally, the real picture can be more complicated, i.e., some hydrocarbons are consumed more than others. Therefore, it can affect the chemistry substantially. In turn, the results of chemical evolution of hydrocarbons determine the composition and structure of forming dust grains. However, in order to model chemistry of hydrocarbons more accurately, it is necessary to know the desorption energy and also the activation barrier of chemisorption for all considered hydrocarbons and different surface structures, which are not available so far.

\subsection{Contribution of cyclotrimerization to the total abundance of aromatics}

To quantitatively assess the contribution of the surface acetylene CTM, we compared the total abundance of aromatic molecules for models with and without CTM by computing their ratio at each radial position ($n_{\rm PAH}({\rm GR+SR})/n_{\rm PAH}({\rm GR})$). The variation in this ratio with distance is presented in Fig.~\ref{sumabund}~(right), where gas and dust abundances are displayed separately. For the model with $E_{\rm d}=3$~eV, the ratio peaks at $\approx$10 near $R=2.2 \; R_{\star}$ before decreasing to $\approx$0.5 both for gas and dust at $3 \; R_{\star}$. At the next computed position, $R= 3.5 \; R_{\star}$, the ratio again rises up to $\approx$4 and remains constant farther. 

The steep fall of the ratio at $R= 3 \; R_{\star}$ is related to the peak efficiency of benzene production in gas-phase reactions that was mentioned above and seen in Fig.~\ref{abund_gas}~(left). Such a behavior means that benzene is formed in the CTM reaction most efficiently relative to gas-phase reactions at the enhanced temperature corresponding to $R=2.2 \;R_{\star}$ and farther, at $R>3 \;R_{\star}$. Specifically, at $R=3\; R_{\star}$ the CTM reaction still plays a crucial role in the benzene formation but gas-phase reactions become the most efficient and provide the major contribution to the abundance of aromatics. 

As was discussed earlier, the parameter $E_{\rm d}$ determines the results of our study. At $E_{\rm d}=2$~eV, the CTM contribution becomes negligible at $R \lesssim 3\; R_{\star}$, primarily due to inefficient sticking of molecules to grains: acetylene molecules do not remain surface-bound long enough to meet another two molecules. At larger distances ($R > 3\; R_{\star}$), at which gas and dust temperatures decrease, surface reactions gain importance. The reduced desorption efficiency enhances the probability of successive acetylene collisions on grain surfaces. Concurrently, gas-phase benzene formation becomes less efficient, further amplifying the relative significance of surface CTM. At $R=5\; R_{\star}$, the ratio $n_{\rm PAH}({\rm GR+SR})/n_{\rm PAH}({\rm GR})$ reaches a value of $\approx 9$.  

In the case in which $E_{\rm d}=4$~eV, CTM plays a negligible role. At $R=2.2\;R_{\star}$, the total aromatic abundance is identical in both models. The abundance of acetylene in gas is reduced by its rapid adsorption, which dominates over desorption. It leads to complete surface coverage by various hydrocarbons along with acetylene. Consequently, when an acetylene molecule adsorbs, the probability of finding sites occupied by \ce{sC2H2} or \ce{sC4H4} for CTM is reduced. The sites with \ce{sC2H2} or \ce{sC4H4} become rapidly covered by other molecules, terminating the CTM sequence. At $R=2.7\;R_{\star}$, the abundance of aromatics predicted with the model with CTM exceeds the abundance predicted by the model without CTM by a factor of~2. However, this enhancement is negligible indeed and is around computational precision as gas-phase abundances of aromatics are insignificant.
 
\section{Discussion}

Let us discuss key physical quantities related to PAHs predicted by our model and compare them with observational estimates from the ISM. Specifically, we shall discuss the PAH number density, the fraction of carbon locked in PAHs, and the fraction of large PAHs.

Observational studies indicate the average PAH abundance in the ISM of approximately $n_{\rm PAH}/n_{\rm H} \sim 3 \cdot 10^{-7}$ \citep{tielens08, dl07}. The total number densities of aromatic species produced in our model are generally consistent with this value. For instance, models with desorption energies of $E_{\rm d} = 2$ and $3$~eV predict $n_{\rm PAH}/n_{\rm gas}$ values on the order of $\sim 10^{-6} - 10^{-5}$ within the dust formation region. The inclusion of the CTM reaction further enhances $n_{\rm PAH}$ by a factor of a few, depending on the specific model parameters.

However, a direct comparison of number densities requires caution due to a significant discrepancy in PAH sizes. The PAHs in the ISM are substantially larger than those in our current model. Studies on PAH photodissociation under UV radiation suggest that only species with more than 40--50 carbon atoms ($N_{\rm C} > 40 - 50$) can survive in the harsh ISM environment \citep{allain96, murga_shiva}. The limit size is even more drastic if we also take into account interaction of PAHs with shock waves~\citep{micelotta10}. Our chemical network does not include pathways to PAHs of such a size, nor does it contain all possible reactions leading to even our largest considered species (pyrene, $N_{\rm C}=16$). Nevertheless, preliminary conclusions can be drawn. The fraction of PAHs with $N_{\rm C} \geq 10$ (i.e., up to pyrene) among all formed aromatics is too small, rarely exceeding $10^{-4}$. Note that the fraction increases by several times if CTM is included in case of $E_{\rm d} = 3$~eV. 

This indicates that the gas-phase growth mechanisms considered here are inefficient. We admit that we underestimate the growth rate. It could be higher by an order of magnitude or so if parcels undergo more thermal pulses than are simulated and more reactions are included up to PAHs of a size comparable to pyrene. But even in this case, the final fraction of large PAHs would be insufficient. Therefore, it is necessary to incorporate alternative growth pathways, such as chemistry on dust grains \citep{merino14, zhao16}, PAH dimerization \citep{webster22}, polymerization of polyynes \citep{krestinin00}, or the partial aromatization of amorphous carbon materials followed by detachment during restructuring events \citep{mark06, jones13}. The conclusion about the necessity to include alternative mechanisms will be tested in the future by expansions of our chemical network.

Given the size disparity, a more relevant comparison may be for the fraction of elemental carbon locked in PAHs ($f_{\rm C}^{\rm PAH}$). \citet{tielens08} estimated this value to be $\sim 3.5 \cdot 10^{-2}$ for the ISM. In our work, $f_{\rm C}^{\rm PAH}$ -- considering PAHs in both gas and dust phases -- varies from $9.20 \cdot 10^{-4}$ to $8.76 \cdot 10^{-2}$ across different models. The values obtained in the dust formation region for models with $E_{\rm d} = 2$ and $3$~eV are comparable to the interstellar value, while the model with $E_{\rm d} = 4$~eV yields a value approximately two orders of magnitude lower. Thus, assuming the carbonaceous material formed can be efficiently incorporated into the ISM and that the proportional allocation of carbon is preserved (though not necessarily the specific molecular structures), our model can account for the observed fraction of carbon locked in PAHs. The small aromatics formed in our simulations would be unlikely to survive travel to the ISM in their original form but could serve as building blocks for larger PAHs or other carbonaceous grains.

\section{Conclusions}

We present a model for the chemical evolution of hydrocarbons in the inner circumstellar envelope of the C-rich AGB star IRC$+$10216. The primary goal of this work is to incorporate the formation of the first aromatic ring on the surface of dust grains and to evaluate its influence on the resulting abundances of aromatic species. We assume the initial grain material is SiC, which can catalyze the CTM of acetylene to form benzene. Furthermore, we propose that the resulting carbonaceous mantle and growing PAHs continue to catalyze CTM on their carbene-like edges.

This model, named BRAHMA, is based on the framework established by \cite{cherchneff12}. We adopted their chemical reaction network and the physical conditions (gas density and temperature) derived from a hydrodynamical model that includes periodic shock waves from stellar pulsations. This network was expanded to include (or update, when possible) reactions with hydrocarbons up to pyrene (\ce{C16H10}). The system of kinetic equations for gas-phase species was extended to include equations for surface species and dust grain populations.

We computed the abundances of aromatic molecules in the gas and dust phases at seven radial positions from $1.2$ to $5\;R_{\star}$. Several scenarios were investigated: models with and without the surface CTM reaction, each with varying desorption energies of 2, 3, and 4~eV. Our modeling results lead to the following conclusions:

\begin{enumerate}
\item A coherent model that couples gas-phase hydrocarbon kinetics with surface chemistry and carbonaceous dust growth significantly alters the predicted abundances of aromatic hydrocarbons. Consequently, this self-consistent approach is essential for obtaining accurate results.

\item Including surface CTM of acetylene increases the total abundance of aromatics by up to an order of magnitude. This process specifically accelerates the formation of biphenyl, phenanthrene, and pyrene. In contrast, the abundances of naphthalene and acenaphthylene decrease, as their formation pathways require acetylene, which is consumed by the efficient surface reactions.

\item The desorption energy of hydrocarbons from the grain surface is a critical parameter. It directly controls the efficiency of CTM by determining the surface residence time of reactants, which in turn regulates the availability of hydrocarbons for subsequent gas-phase reactions. The activation barrier for chemisorption of hydrocarbons onto the surface is equally crucial, as it controls the rate of dust growth and mantle formation.
\end{enumerate}

In summary, our results indicate that the catalytic CTM of acetylene on both SiC and carbonaceous surfaces is a viable pathway in the circumstellar envelopes of C-rich AGB stars, in particular, IRC$+$10216. However, its precise contribution depends on several poorly constrained parameters, most notably desorption and chemisorption energies, which we aim to refine in the future.

\section*{Data availability}
The developed network is publicly available
at https://zenodo.org/records/17347157. In total, it
includes 194 reagents and 636 reactions.

\begin{acknowledgements}
This work was supported by the grant of Russian Science Foundation №24-22-20104, https://rscf.ru/project/24-22-20104.
\end{acknowledgements}

\bibliographystyle{aa}
\bibliography{refs}

\begin{appendix}

\section{Difference between rate coefficients at N$_2$ and H$_2$ atmospheres}
\label{n2}

The observed trends in pressure-dependent rate coefficients can be understood through the lens of third-body energy transfer theory, particularly through the work of \cite{jasper2020}. Jasper systematically studied third-body collision parameters across over 300 organic molecules including alkanes, alcohols, and hydroperoxides with several bath gases (H$_2$, N$_2$, He, Ar) over temperatures ranging from 300 to 2000 K. His trajectory-based approach yielded generalized expressions for collision cross sections ($\sigma$), Lennard-Jones well depths ($\epsilon$), and energy transfer efficiency, often quantified via an effective rotor formalism (N$_{\rm eff}$). Notably, Jasper found that for hydrocarbons, energy transfer is consistently 25--40\% higher for N$_2$ than for H$_2$, depending on temperature and molecular structure. This difference arises because nitrogen, being heavier and more polarizable than hydrogen, has a larger collision cross section and deeper interaction potentials, enabling more effective vibrational energy dissipation during collisions. In contrast, hydrogen’s low mass and weak polarizability result in poorer energy removal efficiency. These findings provide a theoretical benchmark for understanding the role of bath gases in radical recombination reactions.

\begin{figure*}
\centering
\includegraphics[width=0.9 \textwidth]{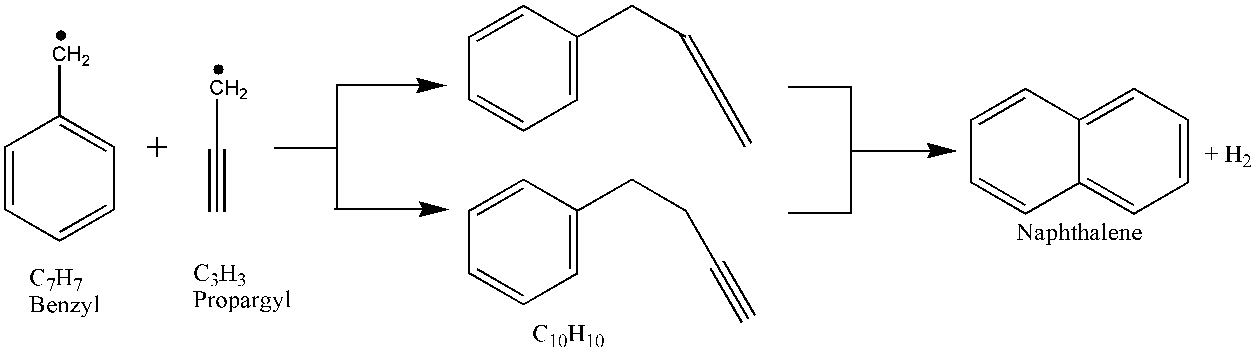}
\caption{Naphthalene formation scheme via reaction of benzyl and propargyl radicals, highlighting initial adducts and final product.} 
\label{napht}
\end{figure*}

Figure~\ref{napht} illustrates the reaction between benzyl (C$_7$H$_7$) and propargyl (C$_3$H$_3$) radicals, which serves as a critical pathway for naphthalene (C$_{10}$H$_8$) formation, a key step in PAH growth. The initial collision of the radicals may also lead to the formation of reactive intermediates such as methyleneindanyl and methyleneindene species. At higher pressures, frequent collisions with the bath gas (H$_2$ or N$_2$) stabilize these intermediates, slowing their progression toward naphthalene. In contrast, at very low pressures, where collisions are rare, the intermediates are less likely to be stabilized, favoring the unimolecular pathway that yields naphthalene and H$_2$ as the final products. The absolute rate coefficients for intermediate formation are up to two orders of magnitude higher than those for naphthalene across the pressure range, but their pressure-dependent trends mirror those of the naphthalene pathway, making the latter a representative case for analysis.

\begin{figure}
\centering
\includegraphics[width=0.41\textwidth]{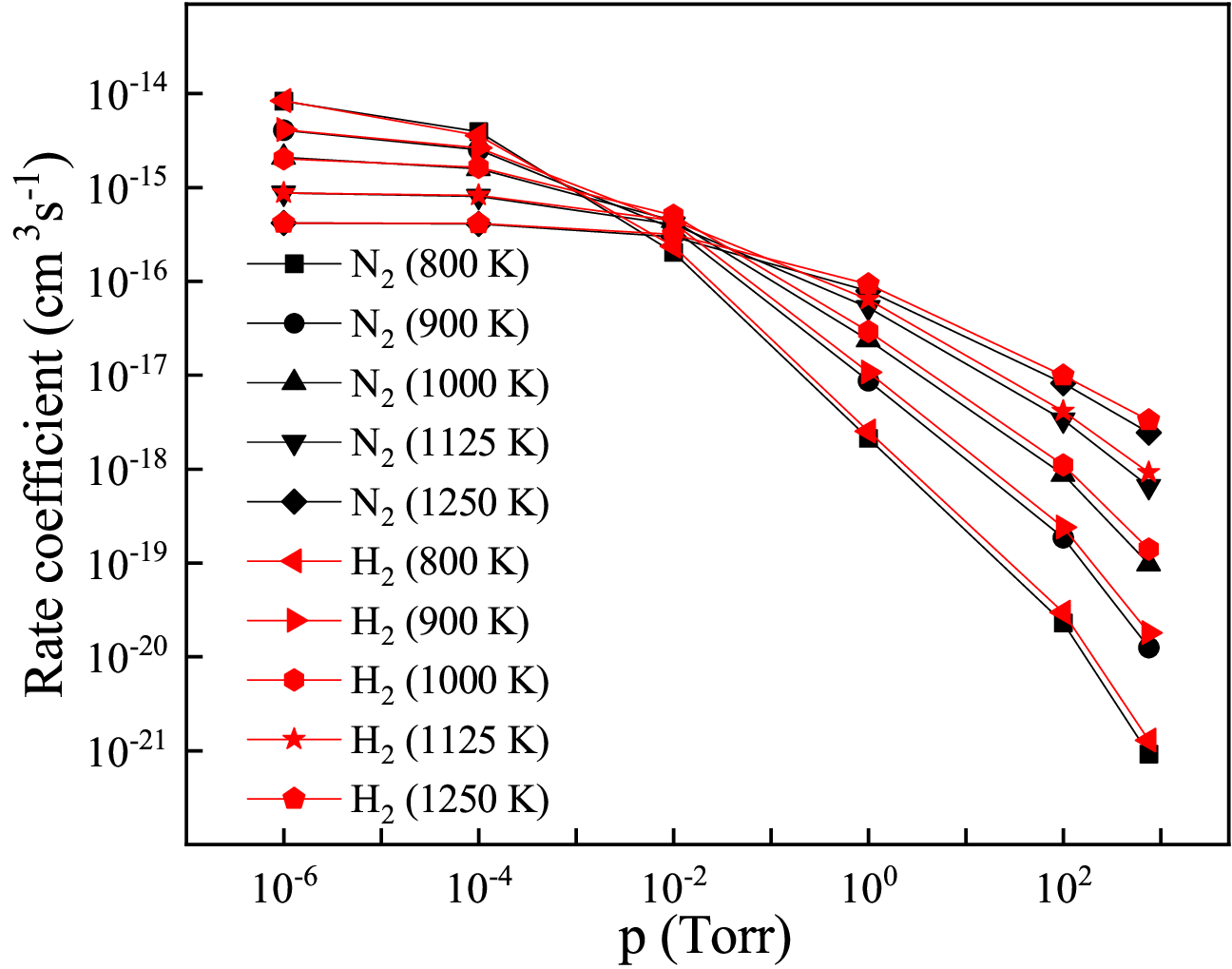}
\caption{Pressure-dependent naphthalene formation rate coefficients at various temperatures, showing convergence at low pressures and divergence at higher pressures.} 
\label{rate-pres}
\end{figure}

The C$_{10}$H$_{10}$ potential energy surface (PES) accessed by the benzyl–propargyl radical reaction has been described in detail in previous publications \citep{matsugi2012,krasnoukhov2022,He2023}. The kinetics and mechanisms of the recombination reaction between benzyl (C$_7$H$_7$) and propargyl (C$_3$H$_3$) radicals have been theoretically investigated using B3LYP, CBS-QB3, and CASPT2 quantum chemical methods. Additionally, a steady-state unimolecular master equation analysis was conducted based on Rice-Ramsperger-Kassel-Marcus (RRKM) theory, with N$_2$ as the bath gas. In this study, we report rate coefficients results using H$_2$ as the bath gas in comparison to N$_2$, employing the same kinetic methodologies used previously. Figure~\ref{rate-pres} presents the pressure-dependent rate coefficients for naphthalene formation at temperatures ranging from 800 to 1250~K. At the lowest pressures ($p<10^{-3}$ Torr), the rate coefficients for both H$_2$ and N$_2$ converge, as the infrequency of collisions becomes the limiting factor, rendering the identity of the bath gas irrelevant. This behavior is typical in astrophysical environments like the ISM, where particle densities are extremely low. As pressure increases into the intermediate range (10$^{-2}$ to 100~Torr), the influence of the third-body properties becomes apparent: N$_2$ consistently yields higher rate coefficients than H$_2$ due to its superior energy transfer efficiency. At even higher pressures, the rate of naphthalene formation decreases further, reflecting the reduced efficiency of the ring-closure pathway under conditions favoring collisional stabilization of intermediates.

The choice of bath gas is context-dependent. In combustion and atmospheric modeling, nitrogen is the standard third body due to its abundance and moderate efficiency. In astrochemical environments, however, H$_2$ dominates as both a collider and a potential reactant, particularly at low pressures where its chemical reactivity can open secondary pathways. This underscores the need for kinetic models that account for both collisional stabilization and chemical reactivity, especially in extreme low-pressure regimes where classical third-body formalisms may fall short.

The pressure-dependent kinetics of the C$_7$H$_7$ + C$_3$H$_3$ reaction not only align with Jasper’s framework for third-body efficiencies but also extend its applicability to a new mechanistic class involving reactive radical intermediates and aromatic ring closure. By corroborating Jasper’s findings, specifically, the superior energy transfer efficiency of N$_2$ over H$_2$, this study bridges his generalized collision parameters to complex PAH-forming reactions. The results highlight how bath gas properties influence not just simple stabilization but also the branching between collisional quenching and chemically driven pathways in radical-aromatic systems, a critical consideration for modeling PAH growth in both combustion and interstellar environments.

\section{Variations in a-\ce{C6H6} abundance with different treatments of dust description}
\label{app: abund_difmod}

\begin{figure}
\centering
\includegraphics[width=0.45\textwidth]{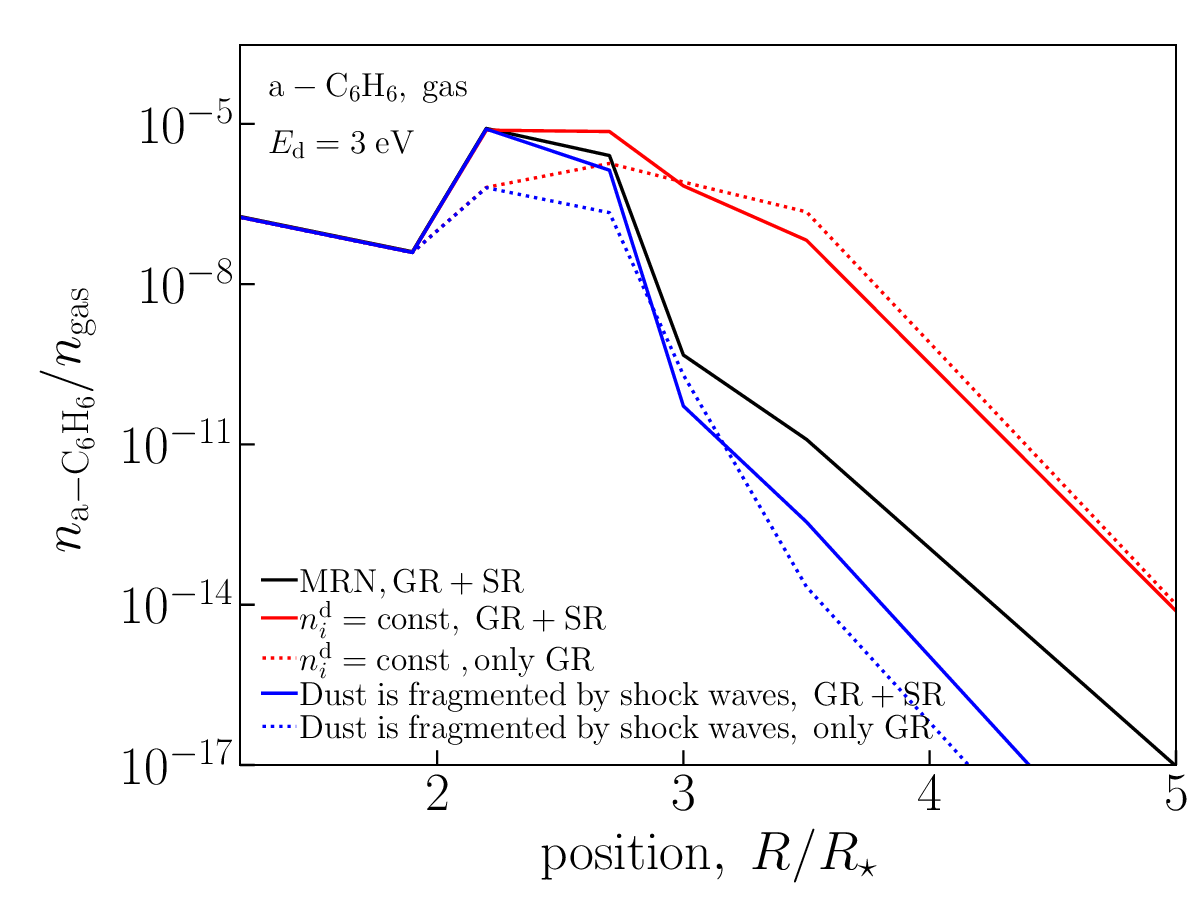}
\caption{Gas-phase benzene abundances at various distances from the star. The black solid line is the result of the ``standard'' model with CMT (Fig.~\ref{abund_gas}). The red solid and dotted lines illustrate the results obtained for the uniform initial size distribution. The blue solid and dashed lines correspond to the cases when dust grains fragment back to the initial size distribution (MRN) at the start of each shock wave.} 
\label{appfig: c6h6_abund}
\end{figure}

In Fig.~\ref{appfig: c6h6_abund} we show the benzene abundance predicted by different versions of the model. Solid lines represent various dust size distributions and demonstrate that the choice of size distribution influences the resulting benzene abundance. Specifically, using a uniform dust size distribution instead of the standard MRN distribution yields a higher benzene abundance. Conversely, if dust grains are assumed to fragment back to the original MRN distribution at the start of each shock wave, the benzene abundance is lower than in models where dust is not fragmented.

Despite these differences, the main conclusion of the paper remains robust: the addition of surface CMT to the reaction network increases the abundance of benzene. This is evidenced by the fact that the solid lines (with CMT) of a certain color are located above the corresponding dotted lines (without CMT), at least within the dust formation region (2–3~$R_{\star}$). Therefore, our findings are not sensitive to the adopted treatment of the dust size distribution.

\section{Dust temperature}
\label{dust_temp}

The dust temperature is calculated from the equation of thermal balance between heating by stellar radiation and due to collisions with gas particles and cooling due to thermal emission (see Sect.~24.1 in \citealt{draine_book} and Sect.~9.2 in \citealt{ivlev_book}):

\begin{eqnarray}
&& 2 \pi \int\limits_0^{\infty} Q_{\rm abs}(\nu, a)  J_{\star}(\nu, R) d\nu + n_{\rm gas} v^{\rm th}_{\rm gas} \gamma k_{\rm B}(T_{\rm gas}-T^{\rm d}) = \nonumber \\ 
&& =  2 \pi\int\limits_0^{\infty} Q_{\rm abs}(\nu, a) B(\nu, T^{\rm d}) d\nu
\end{eqnarray}
where $Q_{\rm abs}(\nu, a)$ is the absorption efficiency for the grain of radius $a$ at frequency $\nu$, $J_{\star}(\nu, R)$ is the intensity of the stellar radiation at frequency $\nu$ and radial distance $R$, $B(\nu, T^{\rm d})$ is the Planck function, $\gamma$ is the thermal accommodation coefficient (adopted to be 0.5 following \citealt{draine_book}). It is assumed that dust grains are mostly heated by collisions with hydrogen molecules, so we do not integrate heating rates over all gas-phase species.

We find $J_{\star}(\nu, R)$ via
\begin{equation}
    J_{\star}(\nu, R) = \frac{1}{2} B(\nu, T_{\star}) \left(1- \sqrt{1-\left(\frac{R_{\star}}{R}\right)^2}\right) \exp(-\tau(\nu, R)),
\end{equation}
where $\tau(\nu,R)$ is the optical depth which is calculated as
\begin{equation}
\tau(\nu,R) = \sum \limits_{i=1}^{N_a} \int \limits_{R_{\star}}^R \pi a_i^2 n_i^{\rm d}(R) n_{\rm gas} Q_{\rm ext}(a_i)dr,
\end{equation}
where $Q_{\rm ext}$ is the extinction efficiency, that is, the sum of absorption and scattering efficiencies. 

\section{Dust size distribution}
\label{app: dust_distr}

\begin{figure}
\centering
\includegraphics[width=0.45\textwidth]{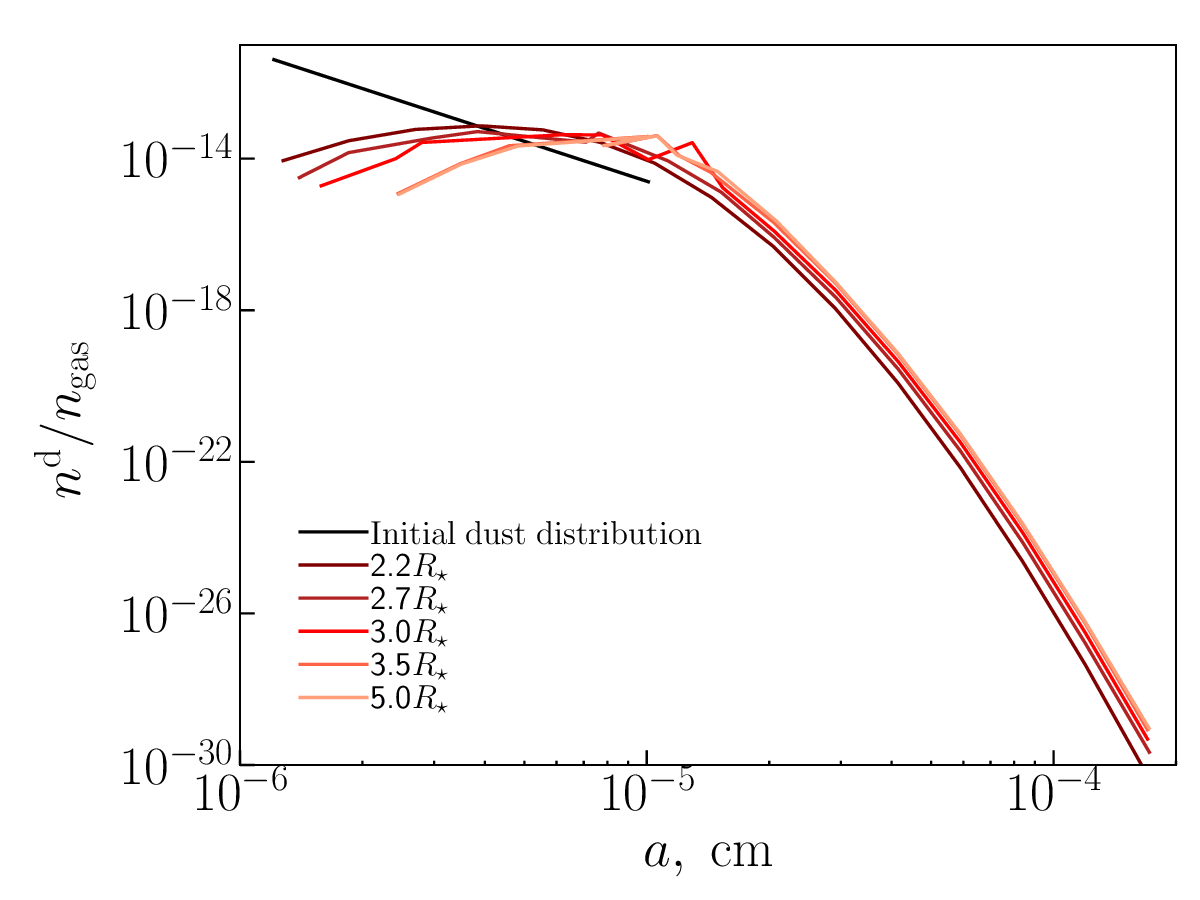}
\caption{Dust size distribution at different positions. The initial distribution is illustrated by black color, colors from brown to orange illustrate the positions indicated in the legend.} 
\label{appfig: dust_distr}
\end{figure}

In Fig.~\ref{appfig: dust_distr} we demonstrate the change of the dust size distribution during evolution of the envelope. Initially there are only grains smaller than 0.1~$\mu$m, and they are distributed by the MRN distribution. Further, dust grains grow via adsorption and coagulation. Consequently, the dust size distribution extends to larger grains, herewith the fraction of small grains decreases. 
 
\section{Calculation results: Extension}

\begin{table*}[h!]
\centering
\caption{Results of the calculations.}
\label{table:summary}

\begin{tabular}{|c|c|c|c|c|c|c|c|c|c|c|}
\hline   
\multirow{2}{*}{$R/R_{\star}$} & \multirow{2}{*}{Model} & \multicolumn{3}{|c|}{$n_{\rm PAH}$} & \multicolumn{2}{|c|}{$n_{\rm C_2H_2}$}   & \multicolumn{3}{|c|}{$f_{\rm C}^{\rm PAH}$} & $f({\rm PAH})$ \\ \cline{3-11}
& & gas &dust& sum & begin & end  & gas & dust & sum & $N_{\rm C}>10$ \\ \hline

\multirow{6}{*}{2.2} &$+$SR, $E_{\rm d}=2$~eV & 8.39e-07 & 7.17e-14& 8.39e-07 & 1.09e-04 & 5.77e-06 & 7.34e-03 & 6.27e-10& 7.34e-03 &5.77e-06\\ \cline{2-11}
&$-$SR, $E_{\rm d}=2$~eV &8.64e-07  & 7.39e-14 &8.64e-07 &  1.12e-04& 9.88e-05 & 7.56e-03  &6.46e-10 &  7.56e-03 & 5.95e-06\\ \cline{2-11}
&$+$SR, $E_{\rm d}=3$~eV & 8.60e-06 & 1.44e-07 & 8.74e-06&  1.10e-04 & 1.24e-05&  7.53e-02 & 1.26e-03 & 7.65e-02 & 1.43e-05 \\ \cline{2-11}
&$-$SR, $E_{\rm d}=3$~eV & 8.16e-07 &1.38e-08 & 8.23e-07 & 1.10e-04 & 9.51e-05 &  7.14e-03 & 1.21e-04 & 7.26e-03 & 5.81e-06\\ \cline{2-11}
&$+$SR, $E_{\rm d}=4$~eV & 3.97e-11 & 1.15e-07& 1.15e-07& 1.09e-04 &7.05e-12& 3.48e-07 &  1.01e-03 & 1.01e-03 & 2.80e-05\\ \cline{2-11}
&$-$SR, $E_{\rm d}=4$~eV & 3.72e-11 & 1.07e-07& 1.07e-07& 1.09e-04& 2.93e-10& 3.26e-07 & 9.37e-04 & 9.37e-04& 2.97e-05 \\ \specialrule{1.5pt}{0pt}{0pt}
 
\multirow{6}{*}{2.7} &$+$SR, $E_{\rm d}=2$~eV & 7.24e-06 & 1.03e-11 &7.24e-06 & 1.02e-04 &8.41e-07 &  6.33e-02& 9.05e-08& 6.33e-02 & 8.41e-07 \\ \cline{2-11}
&$-$SR, $E_{\rm d}=2$~eV & 6.89e-06&  9.83e-12 & 6.89e-06 & 1.05e-04& 7.17e-05 & 6.02e-02& 8.60e-08& 6.02e-02  & 9.31e-07\\ \cline{2-11}
&$+$SR, $E_{\rm d}=3$~eV &  2.93e-06 & 6.94e-06 & 9.86e-06 & 8.30e-05 & 2.65e-08 & 2.56e-02 & 6.07e-02 & 8.63e-02 & 1.39e-05\\ \cline{2-11}
&$-$SR, $E_{\rm d}=3$~eV &  5.63e-07 & 1.45e-06 & 2.01e-06 & 1.02e-04 & 2.47e-05 & 4.91e-03 & 1.27e-02 & 1.76e-02 & 2.69e-06\\ \cline{2-11}
&$+$SR, $E_{\rm d}=4$~eV & 1.30e-11 & 1.07e-07 &  1.07e-07 & 2.47e-07 & 1.83e-12  & 1.13e-07 & 9.36e-04 & 9.36e-04& 2.47e-05\\ \cline{2-11}
&$-$SR, $E_{\rm d}=4$~eV & 1.31e-11 & 1.05e-07 & 1.05e-07 & 2.40e-07 & 1.90e-10 & 1.13e-07 & 9.21e-04& 9.21-04 & 2.47e-05\\ \specialrule{1.5pt}{0pt}{0pt}

\multirow{6}{*}{3.0} &$+$SR, $E_{\rm d}=2$~eV & 5.70e-06 & 3.81e-11 & 5.70e-06 & 6.89e-05 & 1.51e-06 & 4.95e-02 & 3.31e-07& 4.95e-02& 1.51e-06\\ \cline{2-11}
&$-$SR, $E_{\rm d}=2$~eV & 4.60e-06& 3.07e-11 & 4.60e-06 & 7.15e-05& 7.02e-05  & 3.98e-02 & 2.66e-07& 3.98e-02 & 1.94e-06\\ \cline{2-11}
&$+$SR, $E_{\rm d}=3$~eV & 3.40e-08 & 9.70e-07& 1.00e-06 &  2.92e-07 & 2.93e-09  & 2.94e-04 & 8.47e-03 & 8.76e-03 & 1.30e-04\\ \cline{2-11}
&$-$SR, $E_{\rm d}=3$~eV & 2.19e-07 & 3.04e-06& 3.26e-06 & 6.04e-05 & 3.42e-07 & 1.75e-03& 2.55e-02 & 2.73e-02 & 7.07e-07\\ \cline{2-11}
&$+$SR, $E_{\rm d}=4$~eV & 2.93e-12 &  1.05e-07& 1.05e-07 &  2.34e-10 & 2.35e-20 & 2.54e-08 & 9.20e-04 & 9.20e-04 &2.49e-05 \\  \cline{2-11}
&$-$SR, $E_{\rm d}=4$~eV & 2.25e-12 & 1.05e-07& 1.05e-07 & 6.78e-10 & 1.17e-11 & 1.94e-08  & 9.20e-04& 9.20e-04 & 3.25e-05\\ \specialrule{1.5pt}{0pt}{0pt}

\multirow{6}{*}{3.5} &$+$SR, $E_{\rm d}=2$~eV & 3.41e-06& 2.84e-10 & 3.41e-06 & 6.16e-05 & 2.97e-06 & 2.96e-02& 2.46e-06 & 2.96e-02 & 2.97e-06\\ \cline{2-11}
&$-$SR, $E_{\rm d}=2$~eV & 1.76e-06 & 1.47e-10 &  1.76e-06 & 7.05e-05& 7.26e-05  & 1.48e-02 &2.46e-06 & 1.48e-02 &  5.77e-06\\ \cline{2-11}
&$+$SR, $E_{\rm d}=3$~eV & 8.10e-10 & 9.26e-07 & 9.27e-07 & 9.46e-09 & 2.86e-11  & 7.09e-06 & 8.11e-03 & 8.11e-03 & 1.59e-04 \\ \cline{2-11}
&$-$SR, $E_{\rm d}=3$~eV & 1.93e-10 & 2.10e-07 & 2.10e-07 & 6.02e-07 & 1.24e-09 & 1.68e-06 & 1.83e-03 & 1.84e-03 & 2.81e-05\\ \cline{2-11}
&$+$SR, $E_{\rm d}=4$~eV & 6.39e-13 & 1.05e-07& 1.05e-07 & 2.15e-13 & 6.05e-13 & 5.55e-09 & 9.20e-04 & 9.20e-04 & 1.99e-05\\ \cline{2-11}
&$-$SR, $E_{\rm d}=4$~eV & 4.85e-13 & 1.05e-07& 1.05e-07 & 1.19e-11 & 3.96e-12 & 4.22e-09 & 9.20e-04 & 9.20e-04 & 2.66e-05\\ \specialrule{1.5pt}{0pt}{0pt}

\multirow{6}{*}{5.0} &$+$SR, $E_{\rm d}=2$~eV & 3.43e-06  &  7.59e-07 & 4.19e-06 & 3.07e-05 & 2.48e-06 & 3.00e-02 & 6.64e-03 & 3.66e-02 & 2.48e-06\\ \cline{2-11}
&$-$SR, $E_{\rm d}=2$~eV & 3.77e-07 & 8.40e-08 & 4.61e-07 & 7.31e-05 & 6.67e-05 & 3.08e-03 & 6.89e-04& 3.77e-03& 2.24e-05\\ \cline{2-11}
&$+$SR, $E_{\rm d}=3$~eV & 2.80e-12 &  9.27e-07& 9.27e-07 & 3.03e-11 & 2.99e-14 & 2.45e-08 & 8.11e-03 & 8.11e-03 & 2.24e-04\\ \cline{2-11}
&$-$SR, $E_{\rm d}=3$~eV & 6.50e-13 & 2.10e-07 & 2.10e-07 & 1.24e-09 & 1.87e-12  & 5.68e-09 & 1.84e-03 & 1.84e-03 & 4.03e-05 \\ \cline{2-11}
&$+$SR, $E_{\rm d}=4$~eV & 5.21e-16 & 1.05e-07& 1.05e-07 & 3.35e-12 & 7.66e-16  & 4.56e-12 & 9.20e-04 & 9.20e-04 & 2.59e-05\\ \cline{2-11}
&$-$SR, $E_{\rm d}=4$~eV & 5.23e-16 & 1.05e-07& 1.05e-07 & 4.10e-12 & 5.12e-15 & 4.58e-12 & 9.20e-04 & 9.20e-04 & 2.59e-05\\ \hline

\end{tabular}
%\end{center}
\end{table*}

In this section we present Table~\ref{table:summary} with PAH abundances calculated with different versions of our model. The table includes the following columns: 1 -- the distance from the star; 2 -- the model version; 3-5 -- the total PAH abundances in gas, on dust surfaces and summed, respectively; 6-7 -- the gas-phase abundance of acetylene at the beginning and in the
end of propagation of a shock wave; 8-10 -- the fraction of carbon in PAHs in gas, on dust surface and summed, respectively; 11 -- the fraction of PAHs with $N_{\rm C}>10$ among all PAHs. All abundances are given relative to $n_{{\rm gas}}$.

\end{appendix}

\end{document}